\documentclass[aps,twocolumn,superscriptaddress,showpacs,floatfix,longbibliography,nobibnotes]{revtex4-1}
\usepackage{amsmath,amssymb,amsfonts,bm,bbm}
\usepackage{color}
\usepackage[pdftex]{hyperref,graphicx}
\hypersetup{colorlinks = true, urlcolor = blue, linkcolor = blue, citecolor = blue}

\usepackage{amsfonts}
\usepackage{upgreek}
\usepackage{slashed}
\usepackage{latexsym}

\usepackage{todonotes}
\usepackage{mathtools,slashed}
\usepackage[makeroom]{cancel}
\usepackage[normalem]{ulem}
\newcommand{\beq}{\begin{equation}}
\newcommand{\eeq}{\end{equation}}
\newcommand{\ba}{\begin{array}{ccc}}
\newcommand{\ea}{\end{array}}

\newcommand{\str}{\textrm {str}}
\newcommand{\tr}{\textrm {tr}}

\def\bea{\begin{eqnarray}}
\def\eea{\end{eqnarray}}

\begin{document}

\title{Chiral anomaly without Landau levels: From the quantum to the classical regime }

\author{Junhyun Lee}
\affiliation{Department of Physics, Condensed Matter Theory Center and Joint Quantum Institute, University of Maryland, College Park, MD 20742, USA}

\author{J. H. Pixley}
\affiliation{Department of Physics and Astronomy, Center for Materials Theory,
Rutgers University, Piscataway, NJ 08854, USA}

\author{Jay D. Sau}
\affiliation{Department of Physics, Condensed Matter Theory Center and Joint Quantum Institute, University of Maryland, College Park, MD 20742, USA}

\date{\today\\
\vspace{0.4in}}

\begin{abstract}
We study the chiral anomaly in disordered Weyl semimetals, where the broken translational symmetry prevents the direct application of Nielsen and Ninomiya's mechanism and disorder is strong enough that quantum effects are important. 
In the weak disorder regime, there exist rare regions of the random potential where the disorder strength is locally strong, which gives rise to quasilocalized resonances
and 
their effect on the chiral anomaly is unknown. 
We numerically show that these resonant states do not affect the chiral anomaly only in the case of a single Weyl node.
At energies away from the Weyl point, or with strong disorder where one is deep in the diffusive regime, the chiral Landau level itself is not well defined
and the semiclassical treatment is not justified. 
In this limit, we analytically use the supersymmetry method and find that the Chern-Simons term in the effective action which is not present in nontopological systems gives rise to a nonzero average level velocity which implies chiral charge pumping.
We numerically establish that the 
nonzero average level velocity
serves as an
indicator of the chiral anomaly in the diffusive limit. 
\end{abstract}

\maketitle

\newpage

\section{Introduction}

A classical symmetry, which is broken at the quantum mechanical level, introduces an anomaly into quantum field theories. Distinct anomalies appear in different dimensionalities related to the relevant classical symmetry present in each theory. In the context of high-energy physics, such theories must be anomaly free and therefore are canceled in the appropriate construction of the physical problem at hand. However, somewhat surprisingly, such field-theoretic anomalies are ubiquitous in condensed matter systems, with one of the most prominent examples in two dimensions being the parity anomaly and is responsible for the quantum Hall effects. In odd spatial dimensions, when massless Dirac or Weyl fermions are placed in electric $({\bf E})$ and magnetic $({\bf B})$ fields, the axial anomaly is responsible for breaking the charge conservation for each chirality of these massless fermions (in three dimensions this is known as the Adler-Bell-Jackiw anomaly~\cite{Adler69,Bell69}). For an unbounded dispersion, this produces a charge pumping effect (when ${\bf E}$ and ${\bf B}$ are parallel) where one chirality sinks below zero energy and the other chirality rises above, this produces a ``staircase'' of charge moving from one chirality to the other through an infinite Dirac (or Weyl) sea (due to the unbounded dispersion).

The recent discovery of Dirac and Weyl semimetals (e.g., in the compounds $\textrm{Na}_3 \textrm{Bi}~$\cite{Liu14Science,Xu15Science2}, $\textrm{Cd}_3 \textrm{As}_2$~\cite{Borisenko14,Liu14NatMat,Neupane14}, $\textrm{Ta} \textrm{As}$~\cite{Lv15PRX,Lv15NatPhys,Xu15Science1,YangL15}, NbAs~\cite{xu2015discovery}, and TaP~\cite{xu2016observation}) that host linear touching points between the valence and conduction bands at isolated points in the Brilloun zone, represents a unique problem to study the effect of the axial anomaly as the low-energy effective quasiparticles are either massless  Dirac or Weyl fermions. 
Two major distinctions between Dirac and Weyl semimetals and their high-energy ``cousins'' is the fact that they live on a lattice implies both that their energy dispersion relation is bounded (due to the finite size of the Brilloun zone in momentum space) and all Weyl nodes must come in pairs (due to the fermion-doubling theorem~\cite{Nielsen81NPB1,Nielsen81NPB2,Nielsen81PLB}).
This leads to some fundamental distinctions such as the absence of any chiral magnetic effect in equilibrium~\cite{Alavirad16}.
Interestingly, it was predicted that a direct consequence of the axial anomaly in solid-state systems gives rise to a large and negative magnetoresistance in parallel electric and magnetic fields~\cite{Dantas18}.
This was also extended to the semiclassical regime of weak magnetic field strengths~\cite{Son13}. Shortly after the discovery of Dirac and Weyl semimetals, the negative magnetoresistance has now been observed in various compounds such as $\textrm{Na}_3 \textrm{Bi}$~\cite{xiong2015evidence}, $\textrm{Cd}_3 \textrm{As}_2$~\cite{liang2015ultrahigh}, TaAs~\cite{Huang15,Zhang16}, NbAs~\cite{YangX15}, TaP~\cite{Du16}, and NbP~\cite{Shekhar15,Wang16}. 

To construct the appropriate low-energy effective field theory for Dirac and Weyl semimetals that are placed in electric and magnetic fields (required to reveal the chiral anomaly), it is of fundamental importance to incorporate the effects of disorder, which are present in all realistic materials and play no role in high-energy theories.
Disorder can potentially have nonperturbative effects on the dispersion either at energies very close to the Weyl point~\cite{Nandkishore14} or for weak magnetic fields where the scattering rate is higher than the cyclotron frequency.  
At sufficiently low densities, screened charge impurities lead to smooth potentials~\cite{adam2007self,Skinner-2014} that allow us to ignore inter-Weyl-node scattering as the dominant effect.
The two approaches to understand the chiral anomaly so far, i.e., through a chiral Landau level in the absence of disorder~\cite{Nielsen83} (or weak potential scattering~\cite{Goswami15})
and semiclassical approaches via the Karplus-Luttinger velocity term~\cite{Son13} as well as the Boltzmann hydrodynamic approach~\cite{Biswas14}, involve fermionic excitations with relatively well-defined energy and momentum, and a sufficiently small damping rate. 
Therefore, a more profound question to address is whether or not the existence of the anomaly requires the presence of reasonably well-defined energy bands at all (such as the Weyl-type dispersion used for the computation of the chiral anomaly~\cite{Nielsen83}), 
for example, does the axial anomaly persist even if the disorder completely eliminates the chiral Landau level or the Weyl point itself that is characterized by a vanishing density of states? 
This is one of the fundamental questions we aim to answer in this paper.

In the absence of any magnetic field, disordered Dirac and Weyl semimetals have garnered a significant amount of theoretical attention~\cite{Fradkin-1986,Goswami-2011,Kobayashi-2014,Moon14,Nandkishore14,Bitan-2014,*Bitan-2016,Brouwer-2014,Altland-2015,Garttner-2015,Pixley15,Sbierski-2015,Leo-2015,Sergey-2015,Altland16,Bera-2015,Liu-2015,Louvet-2016,PixleyPRX16,PixleyPRB16,Pixley2,Shapourian-2015,Sergey2-2015,Guararie-2017,Pixley17,Sbierski-2017,Wilson17,Wilson18}. This is due to the lack of density of states at the Fermi energy making short-range disorder an irrelevant perturbation (within the renormalization group sense) and it was thought that a disorder-driven itinerant quantum critical point separates the semimetal and the diffusive metal at the Weyl node energy (a review from this perspective~\cite{Syzranov-2016}). However, the presence of short-range disorder introduces nonperturbative rare region effects that have been shown to fill in the low-energy density of states and convert the weakly disordered Weyl (or Dirac) semimetal into a diffusive metal for an infinitesimal amount of disorder, effectively eliminating the Weyl point in a strict sense~\cite{Nandkishore14,PixleyPRX16,Wilson17,Pixley17}.
This converts the semimetal-to-diffusive metal transition into a cross over (dubbed an avoided quantum critical point)~\cite{PixleyPRX16,Pixley2,Wilson17,Guararie-2017}. The fate of such physics in the presence of a magnetic field has been currently unknown, and it is in no way obvious how such nonperturbative rare region effects will affect the presence of the axial anomaly. This is rather interesting as the axial anomaly itself is topological in nature and therefore is a separate nonperturbative phenomenon. 

In this paper, we address both the existence and the indicator of the chiral anomaly in Weyl systems when the conventional Nielsen-Ninomiya's charge pumping mechanism~\cite{Nielsen83} does not (directly) apply. 
There are two main situations in this category.
One is when there are a low density of rare regions in the system which affect the dispersion and especially the chiral Landau level nonperturbatively. 
Due to the drastic change in the energy bands, this case should be investigated separately.
The other is at strong disorder or at finite chemical potential that is away from the Weyl point. Here, disorder should potentially lead to a conventional diffusive metal with regular Ohmic transport. However, for small magnetic fields where the cyclotron frequency is much
 smaller than the scattering rate, the emergence of the chiral anomaly (at a quantum mechanical level) is rather unclear from the traditional field-theory approach. 
This is because one does not expect the formation of any Landau levels, let alone the chiral Landau level. The chiral anomaly in this 
case can be understood as an application of the Karplus-Luttinger velocity to the classical model for transport of quasiparticles~\cite{Son13}. 
Such a classical model cannot describe the chiral anomaly in the strong disorder limit where the scattering rates and Fermi energy become comparable. 
Our main purpose is to study how disorder affects the existence of the chiral anomaly in these cases, and develop a formal comprehensive indicator of the chiral anomaly which can include the two cases as well as generic Dirac and Weyl semimetals. 

The rest of the paper is organized as follows. 
In Sec.~\ref{sec:model} we set up the lattice Hamiltonian and its continuum model that we use throughout the paper. 
We also explain the $\phi$ dispersion, which is the main object we calculate numerically, and how it connects to our central subject of the chiral anomaly. 
Then, in Sec.~\ref{sec:weak} we start our investigation on the chiral anomaly beyond the Nielsen-Ninomiya picture with the numerical study of rare regions and their effect in weak disorder. 
We consider both cases of a single Weyl node and a pair of nodes in this situation.
In Sec.~\ref{sec:strong}, we consider the possibility of a chiral anomaly at an energy away from the Weyl point where intranode scattering should lead to a conventional diffusive metal. Scattering in the diffusive metal is known to eliminate the low-energy fermionic degrees of freedom in favor of diffuson modes of a nonlinear sigma model that naturally describes Ohmic transport. We show that the topological term~\cite{Altland16} in this sigma model leads to a nonzero expectation value of the level velocity as a function of flux that can be shown to lead to the chiral anomaly with a sign that is consistent with the Fermi-surface monopole charge~\cite{Haldane14}, but with a different scaling from the clean case studied previously. 
This is explicitly checked by numerics in the cases we have discussed in Secs.~\ref{sec:weak} and \ref{sec:strong}.
We summarize our results in Sec.~\ref{sec:disc}.

\section{Model}
\label{sec:model}

\subsection{Lattice model for Weyl fermions}
\label{sec:Models}

Weyl materials consist of pairs of linearly dispersing Weyl nodes at low energy. 
To achieve this feature on a lattice in an efficient way, we adapt the following tight-binding Hamiltonian defined on a simple cubic lattice \cite{Wilson17,Wilson18} to describe Weyl fermions:
\begin{align}
	H =& \sum_{i,\eta =x,y,z} t_\eta \, \psi_i^\dagger \sigma_z \psi_{i+\eta} 
		+  i \sum_{i,\alpha = x,y} t'_\alpha \, \psi_i^\dagger \sigma_\alpha \psi_{i+\alpha} 
		+ \textrm{H.c.} \nonumber \\
		&+ \sum_i \psi_i^\dagger (V_i - m \sigma_z) \psi_i .
		\label{eq:Hlatt}
\end{align}
Here, $\psi_i=(c_{i,\uparrow},c_{i,\downarrow})^T$ is a two-component spinor that is composed of fermionic operators $c_{i,\sigma}$ ($c_{i,\sigma}^\dagger$) that are the annihilation (creation) operator at site $i$ with spin $\sigma$; 
$t_\eta$, $t'_\alpha$ are spin-dependent hopping parameters, which we choose $t_x = t_y = t_z = t$ and $t'_x = t'_y = t'$;
$m$ is a constant ``mass''  parameter that controls the location of the Weyl nodes; the $\sigma$'s are the Pauli operators acting in spin space; and $V_i$ is the random disorder potential. 
We choose the random potential to be given by a correlated Gaussian distribution with zero mean and variance of $W^2$: $\langle V_i \rangle = 0$, $\langle V({\bf k}) V(-{\bf k}) \rangle = W^2 e^{-|{\bf k}|^2 / k_0^2}$. 
$k_0$ is the scale of correlation: $k_0 \rightarrow \infty$ gives no correlation and smaller values of $k_0$ indicate stronger correlation. 

We apply a constant magnetic field in the $z$ direction to Eq.~(\ref{eq:Hlatt}), which we include
by Peierls substitution: 
$t_y \mapsto t_y e^{-iBx}$, $t'_y \mapsto t'_y e^{-iBx}$ for all sites, and $t_x \mapsto t_x e^{-iBL_x y}$, $t'_x \mapsto t'_x e^{-iBL_x y}$ for the boundary hopping terms between $x = L_x$ and $1$, where $L_i$ indicates the system size in the $i$ direction 
(Note that we are setting the lattice constant $a=1$, as well as $\hbar = e = 1$.) 
In our gauge choice, the periodic boundary conditions in $x$ and $y$ directions, which will be mentioned shortly, restrict the total magnetic flux through the system to be integer multiples of flux quanta, $\Phi_0 = h/e$.
Since we would want to eliminate any boundary effect and concentrate only on the bulk, we also impose boundary conditions to all three dimensions. 
We choose a gauge defined by $t_z \mapsto t_z e^{-i\phi/L_z}$, which allows us to use periodic boundary conditions in the $x$ and $y$ directions, and a twisted boundary condition in the $z$ direction: $\psi(x,y,z+L_z) = e^{i\phi} \, \psi(x,y,z)$.
The twist can also be understood as a flux through the three-dimensional torus, and in the clean limit where translation symmetry is present ($V_i = 0$) the twist angle $\phi$ has an effect of shifting the crystal momentum by $\phi/L_z$: $k_z \rightarrow k_z + \phi/L_z$. 
In the presence of disorder, translational symmetry is broken and momentum is no longer a good quantum number, nonetheless, we can still probe how low-energy states disperse as a function of the twist, which allows us to access the dispersion in a  ``mini-Brillioun zone.''
The number of Weyl nodes in the system is controlled by the mass $m$: there are four Weyl nodes when $|m/2t| <1$, two when $1<|m/2t| <3$, and zero otherwise~\cite{Wilson17}. 
Throughout the paper, we choose the mass parameter as $1< m/2t <3$ and work with one pair of Weyl nodes at ${\bf k} = (0, 0, \pm \cos^{-1} (m/2t - 2))$. 
In the numerical calculations that follow, we use exact diagonalization to determine the energy eigenvalues and eigenstates of Eq.~(\ref{eq:Hlatt}) 
on a linear system size of $L_x = L_y = L_z = L =20$ (i.e., a volume  $V=L^3$).

In Sec.~\ref{sec:strong} we take the low-energy limit of Eq.~(\ref{eq:Hlatt}) and work with the following continuum model for analytical calculations:
\begin{align}
	H_{\textrm ct} = \pm v {\bf k} \cdot \bm{\sigma} + V({\bf x}).
	\label{eq:Hcontinuum}
\end{align}
The $\pm$ corresponds to the two different chiralities of the Weyl fermions, and $\bf{k}$ is the distance in momentum space measured from the Weyl node, with a Fermi velocity $v$,
and $V({\bf x})$ is the continuum limit of $V_i$.
This linearized Hamiltonian is only valid at sufficiently low energies (compared to the bandwidth of the lowest energy band) and momenta ($|{\bf k}| < \Lambda/v$ where $\Lambda$ is the energy cutoff).

In the following sections, we are interested in studying the effects of disorder on the existence of the chiral anomaly both in systems with a pair of Weyl nodes and a single Weyl node.
To investigate the latter, we need the corresponding single-node Hamiltonian of Eqs.~(\ref{eq:Hlatt}) and (\ref{eq:Hcontinuum}).
A single-node Hamiltonian in the continuum limit is straightforward: we simply take one sign in Eq.~(\ref{eq:Hcontinuum}) and it will describe the physics of the single Weyl node of that chirality.
However, in lattice models this is less trivial due to the fermion-doubling theorem \cite{Nielsen81NPB1,Nielsen81NPB2,Nielsen81PLB}, which enforces the fact that on a lattice (i.e., a bounded momentum space) Weyl nodes always come in pairs.
To circumvent this feature, we add a momentum-dependent potential $U({\bf k})$ to the lattice Hamiltonian~(\ref{eq:Hlatt}), where $U({\bf k}) = 0$ for $0 \leq k_z < \pi$ and $U({\bf k}) = U_0 $ for $\pi \leq k_z < 2\pi$, and concentrate on low energies. 
This potential is artificial and nonlocal, but it effectively shifts the second Weyl point (located between $\pi \leq k_z<2\pi$). 
In this limit, the physics stemming from the second Weyl node is invisible in the low-energy regime.
The shape of the potential $U({\bf k})$ does not necessarily have to be a step function in $k_z$, however, the slope of the potential near the boundaries ($k_z \approx \pi,~2\pi$) dictates the effective energy range where we may assume to be in a single-node limit. 
We use the aforementioned step function with $U_0 = 2t$ in our numerical calculations.

\subsection{$\phi$ dispersion and the chiral anomaly}
\label{sec:phiDisp}

As mentioned in the previous subsection, we will use the phase twist in the $z$ direction ($\phi$) to determine the dispersion of the energy eigenstates in the minizone.
\begin{figure*}
\centering
	\vspace*{1mm}
	\includegraphics*[width=165mm,angle=0]{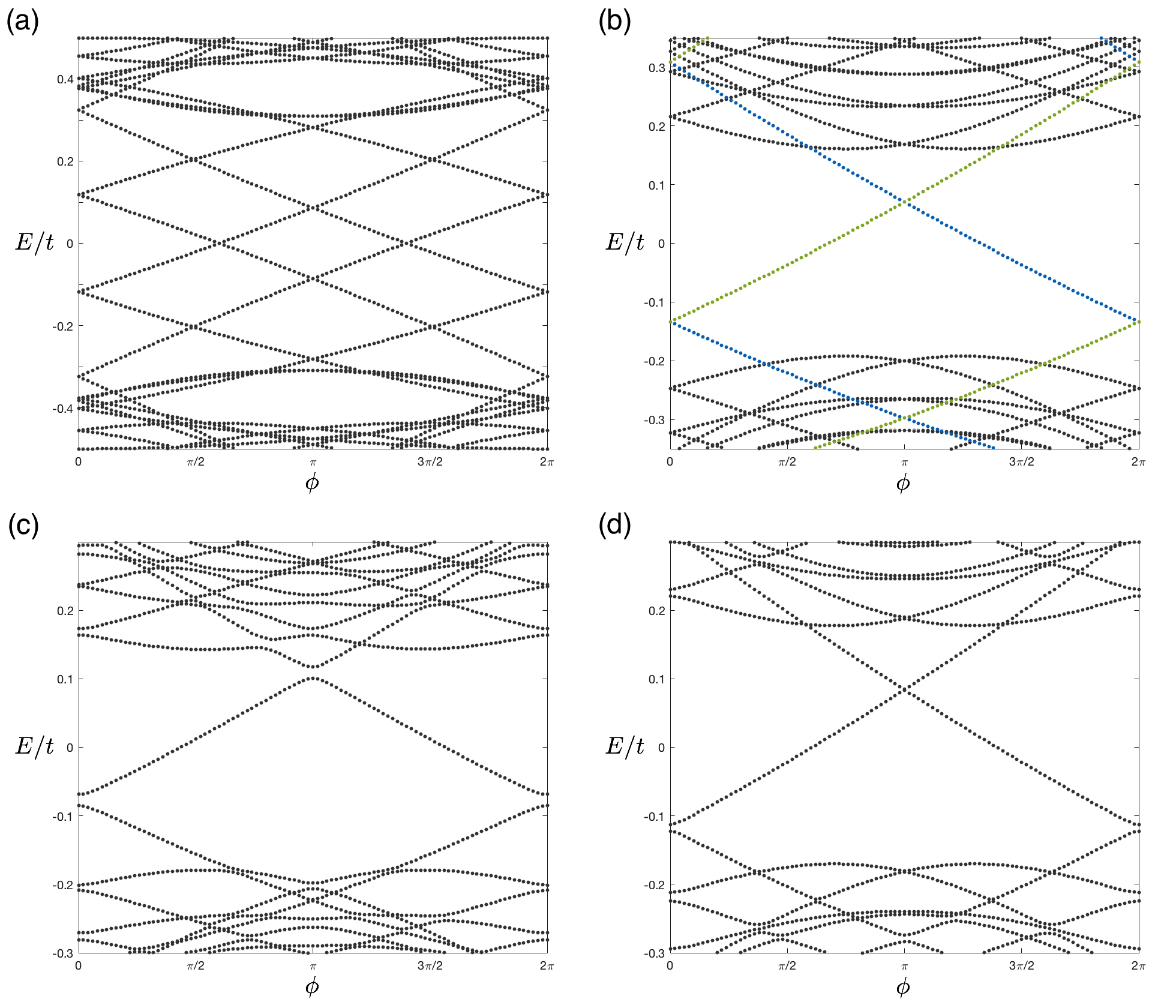}
	\caption{$\phi$ dispersion on the ``minizone'' for several parameters. (a) The clean limit without disorder and also without magnetic field ($W = 0$, $\Phi = 0$). We can observe the two linearly dispersing Weyl nodes. (b) Again the clean limit, but now with one magnetic flux quanta ($W = 0$, $\Phi = \Phi_0$). Landau levels develop due to the magnetic field and the $n=0$ chiral Landau level (CLL) appears. The CLL with positive (negative) chirality is colored in green (blue). (c) The weakly disordered case with one flux quanta ($W = 0.5t$, $\Phi = \Phi_0$). The disorder is uncorrelated ($k_0 \rightarrow \infty$). Due to the uncorrelated disorder the internode scattering is present and thus open gaps when the CLLs with different chiralities intersect (near $\phi = 0,~\pi$). (d) System with weak \emph{correlated} disorder and one flux quanta ($W = 0.5t$, $\Phi = \Phi_0$, $k_0 = k_N/2$; $k_N$ being the $z$ component of the crystal momentum of a Weyl node measured from the $\Gamma$ point). Apart from (c), the disorder correlation suppresses internode scattering, thus significantly reducing the gap between different chirality CLLs. }
	\label{fig:PhiDisp}
\end{figure*}
Figure~\ref{fig:PhiDisp} shows a number of bands in the mininzone from the twist dependence of the eigenvalues of the Hamiltonian~(\ref{eq:Hlatt}) on a $L=20$ cubic lattice.
In the absence of disorder (i.e., $V_i=0$) the states with a phase twist $\phi$ can be interpreted as corresponding to momentum states ${\bf k} = \left(2\pi n_x/L_x,~2\pi n_y/L_y,~(2\pi n_z+ \phi)/L_z \right)$ in the full Brillouin zone, where $n_x,~n_y,~n_z$ are integers. 
This is evident from the clean dispersion in Fig.~\ref{fig:PhiDisp}(a) where the eigenvalue crossings at $E=0$ correspond to states from the Weyl points and the closely spaced states at energies away from 0 result from the quadratically increasing density of states at higher energy.

The application of a magnetic field along the $z$ direction mixes the states between the different momenta $k_x$ and $k_y$ into Landau levels with an index $n$. 
These Landau levels still disperse along $k_z$ as well as $\phi$, similar to Fig.~\ref{fig:PhiDisp}(a). 
However, the $n=0$ Landau level for Weyl nodes is special and disperses with either positive or negative velocity $v(\phi) = dE/d\phi$ depending on the topological charge of the Weyl node. 
This is consistent with the spectrum in Fig.~\ref{fig:PhiDisp}(b), when the system has no disorder and one magnetic flux, where the positive (green) and negative (blue) velocity modes arise from different Weyl nodes. 
Such $n=0$ Landau levels that disperse in a particular direction, carry current only in the same direction and are referred to as chiral Landau levels (CLLs). 
The application of an electric field $\mathcal{E}$ along the $z$ direction leads to an increase in momentum for electrons $\mathcal{E}= L_z^{-1}d\phi/dt$. 
This also results in the Fermi energy shifting according to the equation $dE_F/dt=v\mathcal{E}L_z$, which has a sign that depends on the topological charge of the Weyl point. 
Thus, positively charged Weyl points accumulate charge for parallel electric and magnetic fields, while negatively charged Weyl points lose charge. 
The apparent violation of charge conservation that arises from focusing at a single Weyl node is referred to as the chiral anomaly.

The introduction of finite disorder $V_i$ mixes the states within the full Brillouin zone and can mix different Landau levels leading to the elimination of the CLL picture. 
Figure~\ref{fig:PhiDisp}(c) shows how the $\phi$ dispersion changes when we include both external magnetic flux ($\Phi = \Phi_0$) and uncorrelated Gaussian disorder ($W = 0.5t$ and $k_0 \rightarrow \infty$) to the system of Fig.~\ref{fig:PhiDisp}(a). 
We find that even in the presence of disorder, which potentially hybridizes and destroys the CLLs, two states disperse with $\phi$ (from zero energy) along each of the positive and negative $z$ directions. While the spectrum of these states near zero energy appears similar to CLLs, we will show that disorder changes their character away from zero energy and refer to them as chiral Weyl states (CWS). Even at low energy we see that unlike CLLs, which are twofold degenerate for the case of two flux quanta, the disorder potential breaks the degeneracy of the CWSs, while not affecting their direction of propagation. 
The splitting of these bands becomes more prominent when $W$ is increased.
Note that the dispersion is not symmetric under $\phi \rightarrow 2\pi - \phi$ anymore as in clean systems due to the random potential.
This will also be more prominent in the following figures where $W$ is larger.

As already discussed, the chiral anomaly in the clean case is driven by the sign (or chirality) of the velocity $v(\phi)$ (=$dE/d\phi$) 
of the CLL near zero energy. 
From the discussion in the previous paragraph and Fig.~\ref{fig:PhiDisp} it is evident that the CWSs continue to have the same velocity properties near zero energy as the CLL and therefore have the same chiral anomaly near zero energy. 
However, the situation is less clear for the disordered case for the dispersion of CWSs.
For example, Fig.~\ref{fig:PhiDisp}(b) clearly has two CLLs, one of each chirality with the corresponding positive/negative spectral flow (as $\phi$ varies from 0 to $2\pi$).
In contrast, the two Weyl nodes in Fig.~\ref{fig:PhiDisp}(c) interact via the uncorrelated disorder and hybridize, and as a result a gap opens (due to an avoided level crossing) when the two CLLs cross.
Now, the hybridized band is a mix of positive and negative chiral bands and thus becomes nonchiral.
The spectral flow of these nonchiral bands is zero, which can be explicitly seen by $E(\phi = 0) = E(\phi = 2\pi)$ for every energy band in Fig.~\ref{fig:PhiDisp}(c). 
Therefore, the nonzero spectral flow as a function of $\phi$ is a direct indication for the nontrivial chirality of an energy band, which is a consequence of the chiral anomaly.

The zero spectral flow and disappearance of the chiral anomaly from the avoided crossing of the CWSs in Fig.~\ref{fig:PhiDisp}(c) is caused by impurity scattering between different Weyl nodes. 
Such elimination of the chiral anomaly is trivial in the sense that it invalidates the basic definition of the chiral anomaly in terms of anomalous charge transfer between the different Weyl nodes, which simply breaks the charge conservation at individual Weyl nodes.
Thus, to guarantee a well-defined chiral anomaly, we use a correlated random potential with a finite value of $k_0$ to reduce the size of the scattering matrix elements with a large momentum transfer. 
This suppresses the matrix element by $\sim e^{-|{\bf k}|^2/ 2k_0^2}$ for a momentum transfer of ${\bf k}$. 
In Fig.~\ref{fig:PhiDisp}(d), where we choose all parameters the same as Fig.~\ref{fig:PhiDisp}(c) but $k_0 = k_N/2$, ($k_N$ is the magnitude of the crystal momentum measured from the $\Gamma$ point to the Weyl node in the clean limit) one clearly observes less band mixing between CLLs compared to Fig.~\ref{fig:PhiDisp}(c), indicating the scattering between nodes has been suppressed. 
In the following sections when we want to suppress the internode scattering, we choose $k_0 = k_N /10$, which will suppress the internode scattering matrix element by a factor of $\sim e^{-200}$, and we can assume disorder increases randomness while (almost completely) preserving the topological properties of the Weyl system. 
However, when we want to suppress internode scattering \emph{completely} we consider the model with a single Weyl node.

Considering the spectrum in Figs.~\ref{fig:PhiDisp}(b), (c), and (d) it is clear that CWSs can either undergo avoided level crossings with other states or merge with the continuum and 
the sign of the velocity becomes random, which effectively
flips the direction of their velocity potentially interfering with the chiral anomalous response. 
The chiral anomaly in this case, which is defined as the total rate of charge accumulation in the vicinity of a Weyl point, depends on the velocities of all levels near the Fermi energy. 
In the remaining sections of the paper we will quantify the sense in which the chiral anomaly survives both in the vicinity of the Weyl point and substantially away from the Weyl point when the model is deep  in the diffusive metal regime.
Also note that the effect of the chiral anomaly will only be observable when the chemical potential is within the bandwidth of the CLL in the clean case.
In our numerical calculations due to the single particle nature of the Hamiltonian in Eq.~\eqref{eq:Hlatt}, the chemical potential is set by the energy value under consideration.
Therefore, although we do not set any particular value for the chemical potential throughout the paper, we still assume that it is within the bandwidth of the CLL.

%%%%%
%%%%%
%%%%%

\section{The Chiral anomaly near The Weyl point: rare states}
\label{sec:weak}

Using the framework in the previous section, we study the nontrivial effects of the random potential on the chiral anomaly.
In this section, we concentrate on the effects of weak disorder at low $|E|$, near the Weyl point. 
We define the weak disorder regime where each sample has well-defined low-energy bands in the minizone (as a function of $\phi$), which  are well separated from each other. 
When the magnetic field is present (in the clean limit), the CLL is clearly present in this regime. We will distinguish trivial and chiral bands by studying the spectral flow of  an eigenstate $E$ in one pumping cycle, which is formally captured by $\int_0^{2\pi} d\phi \,\, v_E(\phi) = E(\phi=2\pi) - E(\phi=0)=\delta E$. As we have discussed, trivial bands have $\delta E=0$ while chiral bands have $\delta E \neq 0$, and thus this can be directly observed from the $\phi$ dispersion.

\subsection{Rare states}

In the weak disorder and low $|E|$ limit, the random potential will have two very distinct effects. 
Due to the perturbative irrelevance of disorder (within the renormalization group sense), 
one is the change in energy levels, directly following perturbation theory in the random potential~\cite{PixleyPRX16}. In the presence of the external magnetic field, this also breaks the degeneracies of the Landau levels.
However, a weak broadening (in a disorder averaged sense) of the CLLs essentially only breaks the conserved momentum but does not influence  the conventional spectral flow (Sec.~\ref{sec:phiDisp}) because  the spectral flow through each CLL will remain the same provided no gaps open in the spectrum (which we achieve by either a correlated disorder or a single Weyl node model).
Second, is the effect of the rare regions of the random potential that produces quasilocalized resonances near $E=0$ \cite{Nandkishore14,PixleyPRX16}.
Rare regions produce power-law-localized eigenstates with nonzero level repulsion that are consistent with random matrix theory statistics 
(i.e., are not Anderson localized) and the rare wave function decays at short distances ($r \ll L$) like $\psi(r) \sim 1/r^2$ centered about a rare region (site or cluster of sites) of the potential. 
In the absence of a magnetic field, the rare eigenstates contribute to the low-energy density of states (DOS) and it acquires a nonzero average value at $E=0$: 
$\nu(E)\approx \nu(0) + aE^2$ with $\nu(0)\sim \exp[-(t/W)^2]$ (for an uncorrelated Gaussian disorder distribution) \cite{Nandkishore14,PixleyPRX16}.
Due to the power-law nature of these rare states, samples that have multiple rare regions will have nonzero tunneling matrix elements between them \cite{Nandkishore14,PixleyPRX16}.
In the presence of the magnetic field, as long as the cyclotron orbits are not sufficiently smaller than the quasilocalized wave function, it is natural to expect them to persist (as we will show and see in Appendix~\ref{sec:AA}), however, what their effect will be on the axial anomaly is in no way clear.

We tune the strength of disorder in the proper range which is not too small that the probability of finding a sample with rare region is unrealistically small, nor too large that rare states proliferate the entire system. 
In the numerical calculations in this section, we use $W = 0.7 t$ and search through many different disorder realizations for rare states. 
With this parameter, we were able to find several disorder realizations with rare states among $10 000$ samples.
We first identify the rare state candidates in the absence of a magnetic field by plotting the $\phi$ dispersion in the minizone for a given disorder realization, and check the existence of a nondispersing state \cite{Nandkishore14,PixleyPRX16} near zero energy, that possesses a wave funciton that decays like $1/r^2$ to within numerical accuracy (see Appendix~\ref{sec:AA}). 

%%%

\subsection{Rare states in the single Weyl node model}

\begin{figure*}
\centering
	\vspace*{1mm}
	\includegraphics*[width=165mm,angle=0]{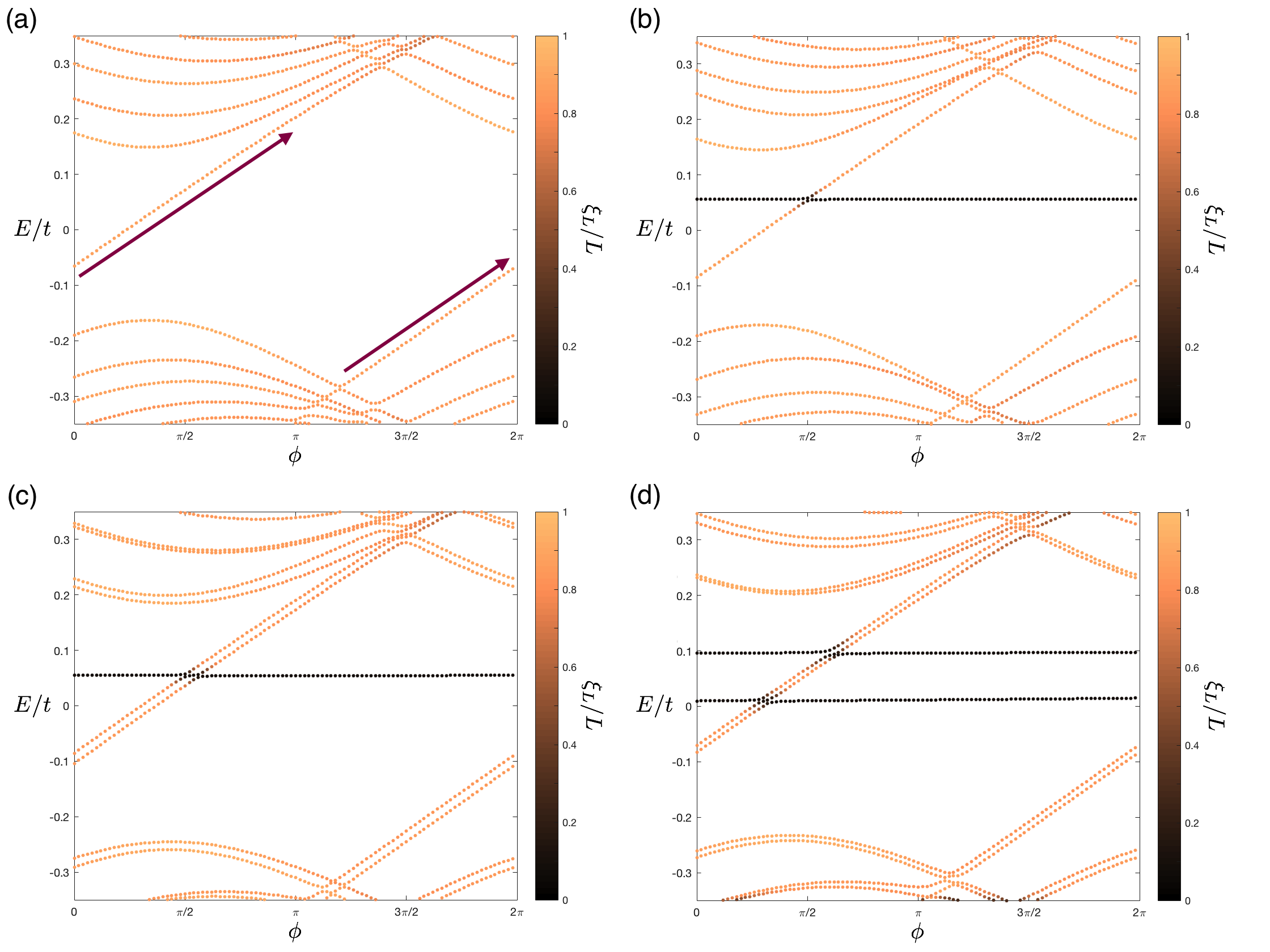}
	\caption{$\phi$ dispersions of the single Weyl node model. 
    All four cases have an uncorrelated disorder potential of strength $W = 0.7t$, but with a different disorder realization. Each state is color coded with their effective IPR localization length $\xi_L$. 
    (a) $\Phi = \Phi_0$ case when the disorder realization does not contain any rare region. The positive chirality CLL, i.e., the linear band at low energy, can be clearly seen and the positive spectral flow is indicated with the arrow. 
    (b) $\Phi = \Phi_0$  when the disorder realization contains one rare region. The flat band with very low $\xi_L$ is the quasilocalized rare state. Notice the hybridization of the rare state and CLL. 
    (c) $\Phi = 2\Phi_0$ case for the same disorder realization as (b). The twofold degeneracy of Landau levels is weakly broken due to the perturbative effect of the disorder potential. 
    (d) $\Phi = 2\Phi_0$ case when the disorder realization contains two rare regions. The two rare states both interact with the CLLs, opening gaps in all four intersections. 
    Note that the total chiral spectral flow is intact despite rare states in all cases (b), (c), and (d).}
	\label{fig:RS}
\end{figure*}
As explained in Sec.~\ref{sec:phiDisp}, the chiral anomaly is closely tied with the spectral flow in the $\phi$ dispersion. 
In this section, we will calculate a number of different $\phi$ dispersions in many conditions both with and without the nonperturbative rare states.
By comparing the spectral flows in the two cases we will be able to determine the influence of nonperturbative effects of disorder on the chiral anomaly in the system.
For conceptual clarity, we first investigate the case with a single Weyl node.
Figure~\ref{fig:RS} shows dispersions of the single-node model 
with
the only difference being the number of external flux [$\Phi = \Phi_0$ for Figs.~\ref{fig:RS}(a) and \ref{fig:RS}(b), and $\Phi = 2 \Phi_0$ for \ref{fig:RS}(c) and \ref{fig:RS}(d)], and the disorder sample [which results in no rare state for Fig.~\ref{fig:RS}(a), one rare state for Figs.~\ref{fig:RS}(b) and \ref{fig:RS}(c), and two rare states for Fig.~\ref{fig:RS}(d); the disorder potential is identical for Figs.~\ref{fig:RS}(b) and \ref{fig:RS}(c)]. 
The extent to which states are localized is quantified through a 
``localization length'' that is defined from the inverse participation ratio (IPR) of each energy ($E$) eigenstate $[\Psi_{{\bf r}\sigma}(E)]$:
\begin{align}
    \xi_L(E) \equiv \left( \sum_{{\bf r}} \, \left [|\Psi_{{\bf r}\uparrow}(E)|^2 + |\Psi_{{\bf r}\downarrow}(E) |^2\right]^2 \right)^{-1/3} .
    \label{eq:xi}
\end{align}
We emphasize that although we use a length scale that measures how localized the states are through the IPR, this does not  imply that the states are exponentially localized.
Note that though the rare states have a relatively short length scale of localization, one should keep in mind that their wave functions decay as a power law and the localization length should be understood strictly in this IPR sense. 

Figure~\ref{fig:RS}(a) is shown as a reference of the spectral flow without any rare states. 
One can clearly observe the positive spectral flow from the CLL, indicating the chiral anomaly is present; the flow is shown as an arrow with positive slope in the figure.
The dispersion in Fig.~\ref{fig:RS}(b) contains one rare state, the nondispersing band near zero energy with small $\xi_L(E)$. 
Since CLLs are extended states, they naturally hybridize with the rare state and open up a gap. 
At first sight, the spectral flow seems to be decreased since $E(\phi = 2\pi) - E(\phi = 0)$ has become smaller due to hybridizing with a rare state. 
However, when considering the spectral flow of the CLL and rare state combined, one can see the \emph{total flow} is the same as that of the CLL in Fig.~\ref{fig:RS}(a). 
We therefore reach one of our main results, namely, the net spectral flow is not affected by a single rare state and the axial anomaly survives nonperturbative effects of disorder for the case of a single Weyl node. 
At energies in the vicinity of the rare states where the spectral flow continues through the rare states, the character of the rare 
states is clearly different from the CLLs and therefore should be considered as part of the broader CWSs.

This feature is not restricted to one flux quanta or single rare state.
Figure~\ref{fig:RS}(c) shows the spectral flow with two flux quanta.
Due to the hybridization of the rare state with both CLLs, the two CLLs are hybridized with each other \emph{through} the rare state. 
But, as in the $\Phi = \Phi_0$ case, the total spectral flow of all three states is unchanged from the existence of the rare state; this can be generalized to multiple fluxes through the system.
Figure~\ref{fig:RS}(d) is when $\Phi = 2\Phi_0$ and with two rare regions in the system. 
It can be seen from the dispersion that basically the same mechanism will hold for samples with multiple rare regions; this may be generalized to cases that have any number of rare regions in the system. 
Combining these observations, we conclude that the total spectral flow is not affected by rare states in any external field, indicating that the chiral anomaly is intact despite the presence of rare states in the single-node model.
 
%%%

\subsection{Rare states in the two Weyl node model}

\begin{figure}
\centering
	\vspace*{1mm}
	\includegraphics*[width=90mm,angle=0]{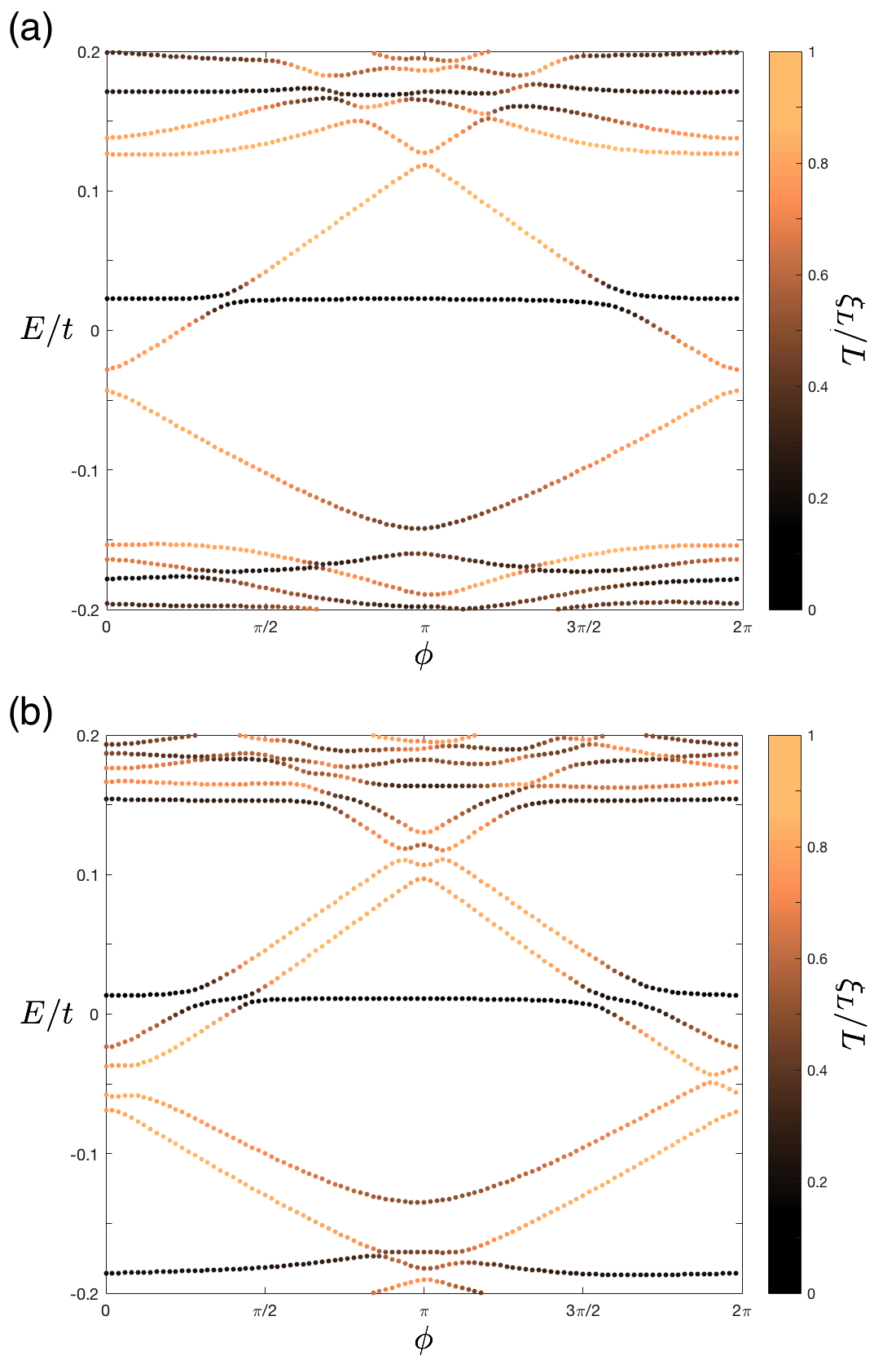}
	\caption{$\phi$ dispersion for rare states with two Weyl nodes. The disorder strength is the same as in Fig.~\ref{fig:RS}, $W = 0.7t$, and the specific disorder realization is chosen as the one with a single rare region. (a) $\Phi = \Phi_0$ case. One can observe the rare state (low $\xi_L$ state around $E/t \approx 0.02$) hybridize with both CLLs. The CLLs loses its chirality after hybridizing, which can also be checked from the zero spectral flow as $\phi$ is increased from $0$ to $2\pi$. (b) $\Phi = 2\Phi_0$ case for the same disorder realization. The rare state mixes with all CLLs and the gap opens in four places. Note the difference from (a) that only one pair (with lower energy) of the CLLs has lost its chirality due to hybridization with the rare state. The other pair of CLLs still remain chiral and survive as a channel of charge pumping. }
	\label{fig:RS2}
\end{figure}
Now, we turn to the physically more realistic case of systems with two Weyl nodes with internode scattering. 
Figure~\ref{fig:RS2} is the $\phi$ dispersion calculated for such systems, with either one or two external flux quanta threading though the system, respectively.
As we have mentioned earlier, the hybridization of CLLs of opposite chirality occurs in the vicinity of high-symmetry points $\phi = 0, \, \pi$ in this model with weak disorder. 
Therefore, by choosing a disorder realization where the rare state band and CLLs cross at $\phi$ far from $0$ or $\pi$ and at a much lower (absolute value of) energy, we can single out how the nonperturbative rare states affect the CLLs and the chiral anomaly separately from typical states that hybridize with each other at much higher energies. 

To be concrete, let us take a look at Fig.~\ref{fig:RS2}(a). 
This shows a particular system where $\Phi = \Phi_0$ and $W = 0.7t$. 
It is apparent from the gaps at $\phi = 0, \, \pi$ that the CLLs have lost their chiral properties. 
However, the rare state which has an energy of $E/t \approx 0.02$ does not cross the CLL on these values of $\phi$.
So, to investigate the rare state effect on the CLLs we only need to consider the regions where the rare state and CLL band meets, i.e., we may concentrate on the energy window of, for example in this case, $ 0 <E/t < 0.05$.  
In the two crossing points where the rare state dispersion intersects with the two CLLs \emph{separately}, we can see that the gap opens at both points. 
These gaps destroy the chiral nature of the CLLs, and the energy band crossing $E = 0$ will not have any spectral flow as we tune $\phi$ from $0$ to $2\pi$ (again, regardless of any possible gap openings in other points of the spectrum). 
In other words, the two different CLLs do not only interact directly, but can also interact via the rare states and lose their topological properties. 
Thus, in this particular case of Fig.~\ref{fig:RS2}(a) the system  does not exhibit the chiral anomaly due to internode scattering, but even assuming that there was no internode scattering (which is in principle possible by increasing the correlation of the disorder), the anomaly would be destroyed by the rare state separately hybridizing with each CLL.

This is a demonstration of one case where the rare state mediates an interaction between the two CLLs, but we can argue that this effect is in fact general. 
Since rare states decay in space as $\psi(r) \sim 1/r^2$, their wave function in momentum space is spread out and can hybridize with (essentially) any plane wave like state~\cite{Pixley17}.
Therefore the rare states generally have wave function overlap with CLLs with both chiralities and thus hybridize with them.
The overlap with each CLL wavefunctions will dictate the size of each gap, which will both be nonzero.
Thus, the argument for Fig.~\ref{fig:RS2}(a) will generally hold for any disorder realization with rare states.

We next consider Fig.~\ref{fig:RS2}(b), the case with the same disorder realization and parameters from Fig.~\ref{fig:RS2}(a) but the external flux is doubled to $\Phi = 2\Phi_0$.
Similar to the case with one flux quantum, we may only concentrate on the small energy window around that of the rare state and ignore all other features of the dispersion.
The rare states cross with two CLLs of each chirality and a gap opens at all four level crossings.
Let us first consider the CLLs with lower energy. 
The situation for the low-energy CLLs is very similar to that of the system with one flux: they hybridize with the rare state and the two opposite chirality CLLs are connected via the rare state.
These are no longer a channel for charge pumping. 
However, the CLLs with higher energy behaves differently. 
They indeed hybridize with the rare state at first, but they again hybridize with the lower-energy CLL.
As a result, the higher-energy CLL does not lose its topological properties and remain chiral, surviving as a channel of the spectral flow.
From this we see that due to the interaction with each rare state, the number of CLLs contributing to the chiral anomaly decreases by one. 

Our observation leads to a conclusion that the effect on the chiral anomaly is modified on a microscopic level. 
Without rare states, the charge imbalance between the two chiralities in the presence of parallel electric and magnetic field is proportional to $\bf{E} \cdot \bf{B}$, which is directly related to the number of flux through the system and the degeneracy of the CLLs.
Now, as per our observation of each rare states eliminating the chirality of one CLL, the pumped charge will be modified by a factor of ``(number of fluxes $-$ number of rare states)/number of fluxes.'' 
In the thermodynamic limit, the number of fluxes scales like $\sim L^2$ whereas the number of rare states scales like $\sim L^3$.
Thus, the rare state effect will dominate the system in the thermodynamic limit except for potentially thin film samples. 

%%%%%
%%%%%
%%%%%

\section{Diffusive topological metal limit}
\label{sec:strong}

Now we turn to the case where the Fermi energy is away from the Weyl point, i.e., the case of the so-called topological metal \cite{Haldane14}, and see whether the chiral anomaly survives in the diffusive limit. 
We follow Haldane in defining the topological metal to be the Weyl semimetal at a Fermi energy sufficiently far from the Weyl point so that the Landau level spacing in a magnetic field is much smaller than the Fermi energy.
The CLL is then only one of the many Landau levels. The addition of weak disorder should lead to a  diffusive metal with a mean-free path that is shorter than the system size so that the DOS becomes independent of the boundary phase twist $\phi$.
The application of a small magnetic field with a magnetic length that is longer than the mean free path leads to a situation where there are no Landau levels and no quantum oscillations of the DOS from the magnetic field. 
One can see this from the fact that quantum mechanically the energy levels do not separate into bands but instead are sufficiently dense and the CLL itself is not well-defined.
Therefore, the spectral flow through a few CWSs, which signals the existence of the chiral anomaly in the low-energy (or clean) limit, is insufficient to describe the chiral anomaly in the diffusive topological metal limit.

The difficulty of defining the chiral anomaly in the diffusive metal limit is resolved by considering the spectral flow of all the states near the Fermi energy, $E$. 
Specifically, the charge added in the vicinity of a specific Weyl point as the phase $\phi$ is incremented  by $2\pi$ is 
\begin{align}
\Delta q&=\int_0^{2\pi} d\phi \, \partial_\phi \sum_j \Theta[E - E_j(\phi)] \nonumber \\
&=-\int_0^{2\pi} d\phi \sum_j v_j(\phi)\, \delta[E - E_j(\phi)],
\end{align}
where $E_j(\phi)$ is the $j$th eigenstate with a twist of $\phi$ and $v_j(\phi) = \partial E_j /\partial \phi$.
Defining the average DOS, $\nu(E,\phi)$, as 
\begin{align}
	 \nu(E,\phi) = \frac{1}{V} \sum_i \delta[E - E_i( \phi)],
     \label{eq:dos}
\end{align}
where $V$ is the volume of the system, we can relate the pumped charge $\Delta q$ to the average velocity 
\begin{align}
&v_{\textrm {avg}}(E,\phi)=\frac{1}{V\nu(E,\phi)}\sum_j v_j(\phi)\,\delta[E-E_j(\phi)]\label{eq:vavg}
\end{align} 
through the equation 
\begin{align}
&\Delta q=-V\int d\phi\,\nu(E,\phi)\,v_{\textrm {avg}}(E,\phi).
\end{align}
In the limit of a system that is much larger than the mean-free path at energy $E$, we expect both the DOS and $v_{\textrm {avg}}$ to be 
independent of $\phi$ so that we obtain a simpler relation 
\begin{align}
&\Delta q=-2\pi V \nu(E)v_{\textrm{avg}}(E).
\end{align}

The average velocity $v_{\textrm{avg}}(E)$ vanishes for most nontopological metals since the spectrum is periodic in the twist $\phi$. Thus, 
the average velocity $v_{\textrm{avg}}(E)$ represents the process of the spectral flow that leads to the chiral anomaly in a single Weyl cone. 
In what follows, we will show that the topological response of the diffusive topological metal indeed leads to a nonzero value of $v_{\textrm{avg}}(E)$ analytically (Sec.~\ref{sec:CCP})
which also agrees with the numerical result (Sec.~\ref{sec:Num}).
We emphasize that this nonzero $v_{\textrm{avg}}(E)$ can serve as a more general indicator of chiral charge pumping and chiral anomaly.

%%%

\subsection{Topological supersymmetric NL$\sigma$M}
\label{sec:SUSY}

We start by reviewing the supersymmetric nonlinear sigma model (NL$\sigma$M)~\cite{Efetov96} approach to determine disorder-averaged spectral properties of noninteracting systems. For a metal, this NL$\sigma$M takes the form of a translationally invariant field theory where the disorder Fermi surface is replaced by a Goldstone mode that describes Ohmic transport. 
%Disordered average properties of non-interacting disordered conductors, including the DOS correlator to appear in Eq.~\eqref{eq:K}, are known to be described by a supersymmetric non-linear sigma model (NL$\sigma$M)~\cite{Efetov96}, which is a translationally invariant field theory where the disorder Fermi surface is replaced by a Goldstone mode that describes ohmic transport. 
Specifically, it is known that the statistical properties at a Fermi energy $E$ of the states of a conventional diffusive metal with a system size larger than the mean-free path $l=v_F\tau$ ($v_F$ is the mean Fermi velocity and $\tau$ is the mean scattering time) is described by the following supersymmetric NL$\sigma$M~\cite{Efetov96,AltshulerPRL93,AltshulerPRB93}:
\begin{align}
	F[Q] = \frac{\pi \nu}{8} \int d{\bf r} \, \str \left[ D\left(\nabla Q - \frac{i e}{c} [Q, {\bf A} \tau_3]\right)^2 + 2i \omega \Lambda Q \right].	
	\label{eq:F0}
\end{align}
$D = v_F^2 \tau/d$ is the diffusion constant, $\bf A$ is the gauge invariant vector potential, and $\omega$ is the frequency difference between the two operators in the correlator we are interested in. 
The microscopic fermion field has been replaced by an $8\times 8$ supermatrix field $Q$;
$\Lambda$ and $\tau_3$ (not to be confused with the mean scattering time) are constant $8\times 8$ supermatrices;
``$\str$'' is the supertrace. 
Details on the fields, parameters, and the NL$\sigma$M itself will follow in Appendix~\ref{sec:AB} and may also be found in Refs.~\cite{Efetov96,AltshulerPRL93,AltshulerPRB93}. 
The NL$\sigma$M, which can be derived from a Hubbard-Stratonovich transformation followed by a gradient expansion about a saddle-point approximation applied to the disordered microscopic Hamiltonian, leads to a microscopic description of Ohmic transport~\cite{AltshulerPRL93,AltshulerPRB93,Efetov96}. 

As we show in Appendix~\ref{sec:AC} (where we closely follow the derivation in Ref.~\cite{Altland16}), from the continuum model for a disordered single Weyl node, 
\begin{align}
	H = v \left( {\bf k} - \frac{e}{c} {\bf a} (\phi)\right)\cdot {\bf \sigma} + V({\bf r}),
	\label{eq:Hcont}
\end{align}
the topological response of the diffusive topological metal~\cite{Haldane14} also appears as a term in the supersymmetric action (a corresponding term also appears in replica analysis~\cite{Altland16}):
\begin{align}
	F_{\textrm{CS}} = \frac{i}{16\pi} \sum_{s = \pm} s \int d{\bf r} \,\epsilon_{ijk} \,\str \left( P^s (\partial_{x_i} A_j) P^s A_k \right).
	\label{eq:CS}
\end{align}
%\begin{align}
%	F_{CS} &= \frac{i}{64\pi} \sum_{s=\pm} s \int d{\bf r} \,\epsilon_{ijk} \,\str \left( (1+ s\Lambda) (\partial_{x_i} A_j) (1+ s\Lambda) A_k \right) \nonumber \\
%    &= -\frac{i B \delta \phi}{16\pi L_z} \int d{\bf r}  \,\str \left( \Lambda Q\right).
%    \label{eq:CS}{}
%\end{align}
%
Here ${\bf a}$ is the vector potential in the system, including that from the external magnetic field and the phase twist, and $P^\pm = (1 \pm \Lambda)/2$ is the projection operator.
In addition to $F_{\textrm{CS}}$, there is an additional topological term in the action that is responsible for the anomalous Hall effect in Weyl semimetals~\cite{Altland16} but since this term does not play a role in the zero mode approximation that we use below, we do not need to consider it here.

We now consider applying the NL$\sigma$M description of disordered Weyl materials reviewed above to compute the level velocity correlator in a magnetic field. As discussed in previous works computing the level velocity correlator~\cite{AltshulerPRB93,Altland16}, in the diffusive regime such correlators can be reasonably determined from the zero-mode approximation of the field theory. The definition of the level velocity depends on the imposition of a phase twist $(\Lambda \delta \phi/ L_z \hat{z})$ across the system. Furthermore, the magnetic field is introduced into the theory through a vector potential that satisfies $\nabla \times {\bf a} = B \hat{z}$ in Eq.~\eqref{eq:Hcont}.
The combination of the vector potential from the magnetic field and the phase twist leads to an interesting contribution from the Chern-Simons term [Eq.~\eqref{eq:CS}], which is proportional to $\str(\Lambda Q)$. 
Following these transformations, the action within the zero-mode approximation for a Weyl metal (including the Chern-Simons term) is written as:
%\mod{Now we use the zero-mode approximation~\cite{AltshulerPRB93,Altland16} and substitute the vector potential as $\nabla \times {\bf a} = B \hat{z}$ with an additional curl-free term $(\Lambda \delta \phi/ L_z \hat{z})$ for the phase twist.
%From this, the Chren-Simons action reduces to a term proportional to $\str(\Lambda Q)$. (See Appendix~\ref{sec:AC}) }
%The \mod{zero-dimentional} final action for topological metals, including the relevent Chern-Simons term is:
%
\begin{align}
	F_0[Q] = \frac{\pi}{8 \Delta} \str \left[ -D \left(\frac{e}{c} [Q, {\bf A} \tau_3]\right)^2 +2i \left( \omega - \frac{B \delta \phi}{4 \pi^2 L_z \nu}  \right) \Lambda Q \right] 
	\label{eq:FQ}
\end{align}
%
%\begin{align}
%	F_0[Q] = \frac{\pi \nu}{8} \int d{\bf r} \, \str &\left[ D\left(\nabla Q - \frac{i e}{c} [Q, {\bf A} \tau_3]\right)^2 \right. \nonumber \\ 
%     &\left. ~+2i \left( \omega - \frac{B \delta \phi}{4 \pi^2 L_z \nu}  \right) \Lambda Q \vphantom{\left(\frac{i e}{c}\right)^2} \right].
%	\label{eq:FQ}
%\end{align}
%
Here, $\Delta = 1/(V \nu)$ is the mean level spacing.
An important feature of Eq.~(\ref{eq:FQ}) is that the actions with $B \neq 0$ and $B = 0$ are connected by a simple shift in frequency $\omega\rightarrow \omega-\frac{B \delta \phi}{4 \pi^2 L_z \nu}$.
This action will be the key ingredient in determining the topological contribution to the level velocity.
%This is essentially identical to the action at $B=0$ except for a shift of the frequency $\omega\rightarrow \omega-\frac{B \delta \phi}{4 \pi^2 L_z \nu}$.

%%%

\subsection{Chiral charge pumping}
\label{sec:CCP}

We can express the average velocity ($v_{\textrm{avg}}$) in terms of the DOS correlator, 
\begin{align}
    K(\omega,\delta\phi) &= \left\langle  \nu(E, \phi) \, \nu (E+\omega, \phi+\delta \phi)  \right\rangle,
    \label{eq:K}
\end{align}
where the DOS operator is defined as in Eq.~\eqref{eq:dos},
$\langle \,\cdots \rangle$ indicates an average over a range of energy, phase twist, and disorder realizations.
As mentioned earlier, this correlator can be calculated using the NL$\sigma$M introduced in Sec.~\ref{sec:SUSY}.
The mean level velocity is now written as:
\begin{align}
&~v_{\textrm {avg}}(E,\phi)\langle\nu(E,\phi)\rangle \nonumber \\
=&\lim_{\delta\phi\rightarrow 0} \frac{1}{V}\sum_j \left\langle \frac{E_j(\phi+\delta \phi)-E_j(\phi)}{\delta\phi}\delta[E-E_j(\phi)]\right\rangle\nonumber\\
=&\lim_{\delta\phi\rightarrow 0}V\int_{\Delta\omega}d\omega\,\frac{\omega}{\delta\phi}\,K(\omega,\delta\phi). \label{eq:vavg2}
\end{align} 
Note that Eq.~\eqref{eq:vavg2} assumes that $\Delta\omega$, which represents a range for the integration of $\omega$, to be an amount that is much smaller than the mean level spacing [$\Delta = 1/(V \nu)$] so that the double sum implicit in Eq.~\eqref{eq:K} can be approximated by the single sum over states in Eq.~\eqref{eq:vavg2}.

Now, we can calculate $v_{\textrm{avg}}(E,\phi)$ for the single Weyl node Hamiltonian~\eqref{eq:Hcont}, which results in the NL$\sigma$M of Eq.~\eqref{eq:FQ}.
Motivated by the shift of frequency in Eq.~\eqref{eq:FQ} eliminates its $B$ dependence, it is convenient to shift $\omega\rightarrow \omega+\frac{B \delta \phi}{4 \pi^2 L_z \nu}$. 
Applying this, $v_{\textrm{avg}}(E,\phi)$ takes the form
\begin{align}
&~v_{\textrm{avg}}(E,\phi)\langle\nu(E,\phi)\rangle \nonumber \\
=& \lim_{\delta\phi\rightarrow 0}V\int_{\Delta\omega}d\omega\left[\frac{\omega}{\delta\phi}+\frac{B }{4 \pi^2 L_z \nu}\right]K(\omega+\frac{B \delta \phi}{4 \pi^2 L_z \nu},\delta\phi) .
\label{eq:vaveTop}
\end{align}
Considering the $\omega$ dependence of Eq.~\eqref{eq:FQ}, one can observe that 
$K(\omega+\frac{B \delta \phi}{4 \pi^2 L_z \nu},\delta\phi) = K_0(\omega,\delta\phi)$,
where $K_0$ is the correlation function at $B=0$ [i.e., with the $B$ dependence of the frequency term in Eq.~\eqref{eq:FQ} canceled] and also the NL$\sigma$M for a nontopological system. 
Furthermore, the $\frac{\omega}{\delta \phi}$ contribution to Eq.~\eqref{eq:vaveTop}, which is identical to the level velocity of the nontopological system, must vanish so that
\begin{align}
v_{\textrm{avg}}(E,\phi)\langle\nu(E,\phi)\rangle 
= \frac{B }{4 \pi^2 L_z \nu}\lim_{\delta\phi\rightarrow 0}V\int_{\Delta\omega}d\omega K_0(\omega,\delta\phi).
\end{align}
Using the same assumption of small $\Delta\omega$ as in Eq.~\eqref{eq:vavg2}, we can calculate the integral as well as the limit $\delta\phi\rightarrow 0$ which ensures that the $\lim_{\delta\phi\rightarrow 0}\int_{\Delta\omega}d\omega K_0(\omega,\delta\phi)=V^{-2}\sum_i \langle \delta(E-E_i(\phi))\rangle=V^{-1}\langle \nu(E,\phi)\rangle$.
This (and also restoring the units) leads us to the key result of this section:
\begin{align}
	\int_0^{2\pi} d\phi \, v_{\textrm{avg}}(E,\phi) &= \frac{B}{2\pi L_z \nu (\hbar / e)} \nonumber \\
	&= \frac{\Phi}{h/e} \Delta.
	\label{eq:Vphi}
\end{align}
Here $\Phi \equiv B L_x L_y$ is the total flux though the system, and $\Delta = 1/(V \nu)$ is the mean level spacing. 
The above relation suggests that the integration of the mean level velocity from $0$ to $2\pi$, which is directly connected to the chiral charge pumping, is a product of (i) the number of flux quanta through the system, and (ii) the mean level spacing. 

%%%

\subsection{Numerical results}
\label{sec:Num}

\begin{figure*}
\centering
	\vspace*{1mm}
	\includegraphics*[width=165mm,angle=0]{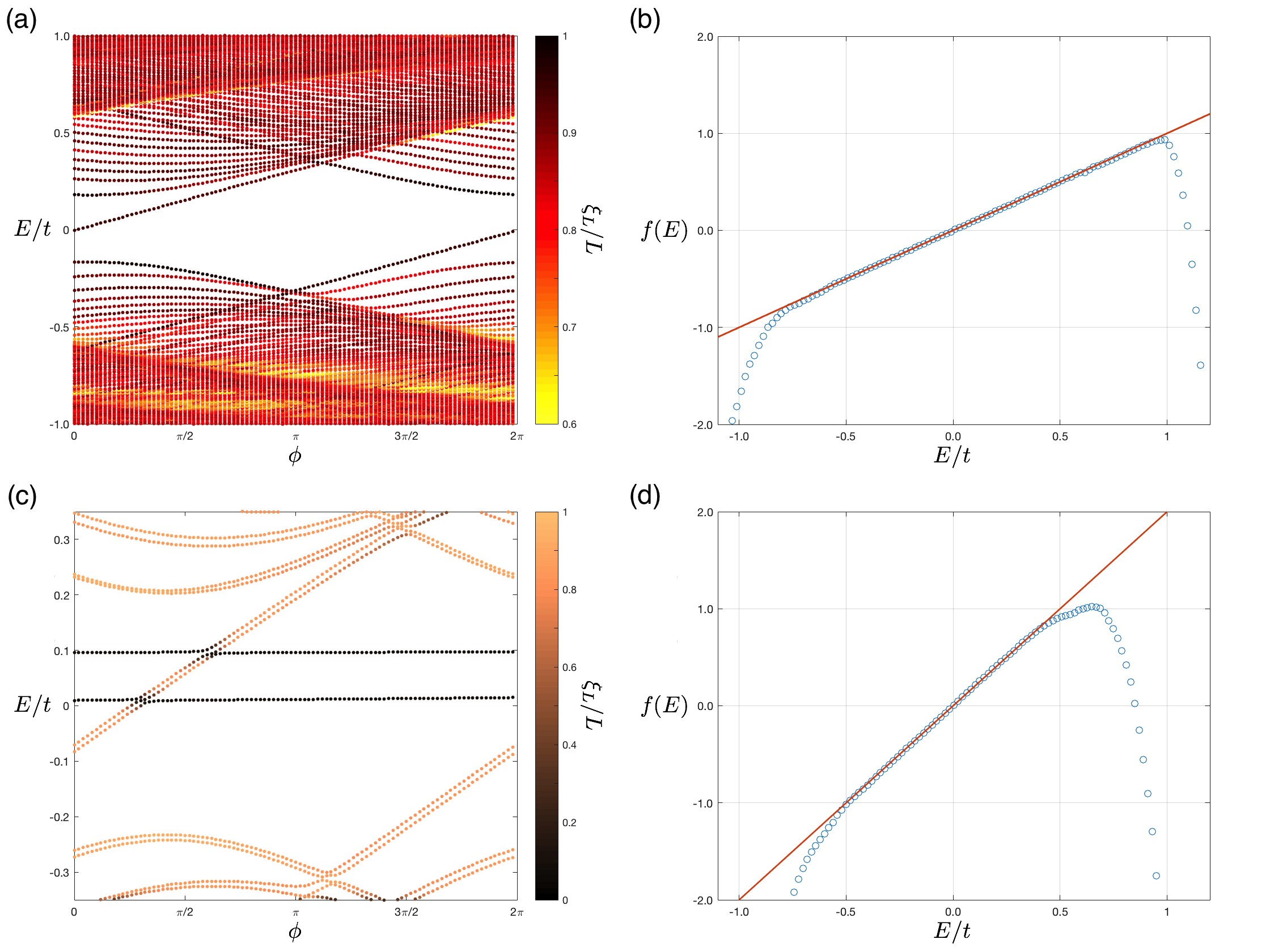}
	\caption{Numerical data on verifying the relation \eqref{eq:fE}.  
    (a) $\phi$ dispersion of a system with parameters $W = 6.0t$, $k_0 = k_N/10$, and $\Phi = \Phi_0$. Notice in the energy range of $|E/t| > 0.5$, the DOS does not have any structure in $\phi$ and can be considered as a topological metal in the diffusive limit. 
    (b) The $f(E)$ function, Eq.~\eqref{eq:fE}, for the system in (a). The data are linear in energy with the coefficient of one, the number of flux, and this feature persists to the diffusive metal limit ($|E/t| > 0.5$). The red solid line is the guide to the eye with slope of one. 
    (c) The same $\phi$ dispersion as in Fig.~\ref{fig:RS}(d), where there are rare states near the Weyl point and $\Phi = 2\Phi_0$. 
    (d) The $f(E)$ function for the system in (c). The data match well that it is linear with the coefficient of two. The red solid line is the guide to the eye with slope of two. 
    }
	\label{fig:Vf}
\end{figure*}
We now analyze our numerical data in light of the derivation in the previous section.
The result for the average velocity of an ideal single Weyl node [Eq.~(\ref{eq:Vphi})] can be confirmed in our numerical calculations.
Since our calculation of the $\phi$ dispersion includes diagonalizing the tight-binding Hamiltonian, we have all the eigenstates of the system together with their corresponding eigenvalues.
To obtain the information of the level velocity, we use the Hellman-Feynman theorem and calculate the expectation value of $\partial_\phi H$, where $H$ is the tight-binding Hamiltonian in Eq.~(\ref{eq:Hlatt}): $v_{E_i} (\phi) = \langle E_i(\phi)| \partial_\phi H | E_i(\phi) \rangle$.
Here, the subscript in the right-hand side indicates expectation value for the state $i$ with twist phase $\phi$. 
Note that the $\phi$ dependence of Eq.~(\ref{eq:Hlatt}) is implicit in the equation for the sake of clarity. 

Now, we define a function $f(E)$ which sums the level velocity for all states in the $\phi$ dispersion which have energy less than $E$.
Also, as per Eq.~(\ref{eq:Vphi}), $f(E)$ should be linear in energy with the slope being the number of flux quanta when $E$ is within the CLL bandwidth.
Combining these we can write:
\begin{align}
	f(E) &= \sum_{E_i \leq E} \, \int_0^{2\pi} d\phi \,\, v_{E_i} (\phi) \nonumber \\
	& = \frac{\Phi}{h/e} (E - E_0) .
	\label{eq:fE}
\end{align}
Here, $E_0$ is the onset energy of $f(E)$, which corresponds to the lowest energy of the CLL in the clean limit. 
Note that $f(E)$ is not a disorder-averaged quantity, and the mean level spacing in Eq.~(\ref{eq:Vphi}) is replaced by the actual energy difference of the particular disorder realization, i.e., dividing Eq.~(\ref{eq:fE}) by the number of states (between energy $E$ and $E_0$) and disorder averaging both sides of the equation leads to Eq.~(\ref{eq:Vphi}).
This equation only holds for a single Weyl node system since it is a direct consequence of Eq.~(\ref{eq:Vphi}).
When we calculate $f(E)$ for the system with two Weyl nodes (of opposite chirality), the contributions from each node will cancel and give zero exactly. 
Lastly, the quantity $f(E)$ is well defined for lattice models, but it can be made finite even for continuum models by introducing a lower bound on the energy sum. Such a lower bound is not expected to qualitatively change the result since $f(E)$ is an averaged quantity.
%whereas within the field theory formulation it will be affected by the regularization.}

Figures~\ref{fig:Vf}(a) and \ref{fig:Vf}(b) show a numerical calculation of the $\phi$ dispersion and a plot of its corresponding $f(E)$ function. 
The system has $W/t = 6.0$, $\Phi = \Phi_0$, and $k_0 = k_N/10$. 
For this case, we find $f(E)$ has a slope of $\Phi / \Phi_0 = 1$ for an energy range near $E = 0$, which is consistent with Eq.~(\ref{eq:fE}). (The red line is a guide to the eye, which has exactly a slope of one.)
It is important to observe that the energy window satisfying Eq.~(\ref{eq:fE}) is not limited to the very vicinity of $E=0$ where states are relatively sparse (in other words, the CLL is relatively apparent), but extends deep into the ``diffusive'' region where the $\phi$ dispersion shows no evident structure. 
 
Note that the overall profile of $f(E)$ is not of the form we have expected, as we can see from the deviation of the data from the red linear line.
However, this is an artifact of the particular way we have constructed the single Weyl node lattice Hamiltonian (Sec.~\ref{sec:Models}). We note that there is also an arbitrary shift in the $y$ axis to make the data pass through the origin. 
The energy range which appears to be linear in the $f(E)$ plot can also be viewed as where the effective single-node approximation of the lattice Hamiltonian is valid.

The field-theory calculations in the previous subsections strictly apply in the limit of the topological metal, where the Fermi energy is away from the Weyl point and writing the NL$\sigma$M is justified. 
However, we find that the derived average level velocity and its dependence of flux [Eq.~\eqref{eq:Vphi}] also holds in the presence of rare regions discussed in Sec.~\ref{sec:weak}. 
Figure~\ref{fig:Vf}(c) is the same $\phi$ dispersion as in Fig.~\ref{fig:RS}(d), and Fig.~\ref{fig:Vf}(d) is the corresponding $f(E)$ of the system.
The $f(E)$ now has a slope of $\Phi / \Phi_0 = 2$ (which is the slope of the red solid line) for the energy range including the two rare states.  
This shows that the level velocity can be a good indicator of the chiral anomaly near \emph{and} away from the Weyl point. Here, the data fall off of the linear energy dependence at smaller energies than in Fig.~\ref{fig:RS}(b) because we have not suppressed intervalley scattering in this sample and at these energies the two Weyl nodes (at different energies) begin to scatter more strongly.

%%%%%
%%%%%
%%%%%

\section{Discussion}\label{sec:disc}
We have shown the existence of the chiral anomaly in two situations where disorder leads to strong violations of 
being able to define sharp 
energy bands in momentum space. In the first situation involving the rare states, disorder leads to quasilocalized states that have no well-defined momentum 
so that they contribute to a featureless continuum in the spectral function~\cite{Pixley17}. Understanding the 
chiral anomaly in this state requires pumping of electrons through states which are hybrids of the chiral Landau level and 
the rare states.  The second situation we have studied  involves a diffusive metal with strong 
disorder. One way this can occur is when the Fermi level is away from the Weyl point and the 
magnetic field is small enough so that the cyclotron frequency is smaller than the disorder scattering rate. The chiral anomaly here, 
despite the absence of chiral Landau levels, has previously been understood semiclassically through the Karplus-Luttinger term~\cite{Son13}
or even with interaction using hydrodynamics~\cite{Biswas14}. 
In contrast, both the field theory and numerical results described in this paper are completely quantum mechanical and 
therefore capable of describing quantum interference effects that are needed to describe the competition between localization 
and the chiral anomaly that forbids localization~\cite{Ryu07,Fu12,Foster12,Fulga14}.
Another way the second situation happens is when a diffusive metal arises from strong disorder near the Weyl point. The semiclassical 
approach to the chiral anomaly~\cite{Son13} is not applicable to this situation because the quasiparticle scattering rate is larger than the 
Fermi energy (which approaches zero). 
However, our analysis continues to apply since our field-theoretic results do not depend on the existence of well-defined energy bands.  The chiral anomaly in disordered Weyl semimetals has previously been described using 
the replica sigma model~\cite{Altland16}. 
%\mod{In such theories, }regularization can sometimes lead to the appearance of a non-vanishing current at vanishing electric field~\cite{Altland16,Goswami-2013}, which is known to be unphysical in real materials~\cite{Vazifeh13,Ma15,Zhong16,Alavirad16,Zhou13}.  
%\mod{Our work, in contrast, avoids regularization issue at all by concentrating on energies near the chemical potential.
%It was still sufficient to find a general indicator of chiral charge pumping as the average level velocity. }
In these works, the conserved charge current is %the key ingredient that 
computed to determine transport properties. 
The charge current in a single Weyl cone receives contributions from energies that are arbitrarily below the Fermi energy and are therefore a quantity that technically depends on the regularization used. 
In fact, most natural regularization schemes in these materials lead to the appearance of a finite current at vanishing electric field~\cite{Altland16,Goswami-2013}, which is known to be unphysical in real materials~\cite{Vazifeh13,Ma15,Zhong16,Alavirad16,Zhou13}. 
In contrast, this work focuses on the characterization of the anomaly in terms of the level velocity correlator near the Fermi energy. 
Since our indicator in Eq.~\eqref{eq:Vphi} depends only on states near the Fermi energy, the quantity we compute has no direct dependence on the regularization [although the action Eq.~\eqref{eq:F0} is derived with the same regularization used in Ref.~\cite{Altland16}]. 
In fact, we expect that our approach of computing the effects of the anomaly through level velocity correlators might be a direct way to study anomalies in disordered topological systems in other symmetry classes.
Finally, the quantum description of the chiral anomaly in strongly disordered system likely paves the way for an understanding of how the 
chiral anomaly preempts Anderson localization.

\acknowledgments{ 
We thank W. Cole, P. Goswami, D. Huse, and S. Parameswaran for insightful discussions.
J.S. acknowledges support from the
JQI-NSF-PFC, the National Science Foundation NSF Grant No. 
DMR-1555135 (CAREER) and the Sloan fellowship program. We acknowledge the University of Maryland supercomputing resources (http://www.it.umd.edu/hpcc) made available in conducting the research reported in this paper.}

%%%%%
%%%%%
%%%%%

\appendix

\section{Rare state wave function}
\label{sec:AA}

\begin{figure}
\centering
	\vspace*{1mm}
	\includegraphics*[width=90mm,angle=0]{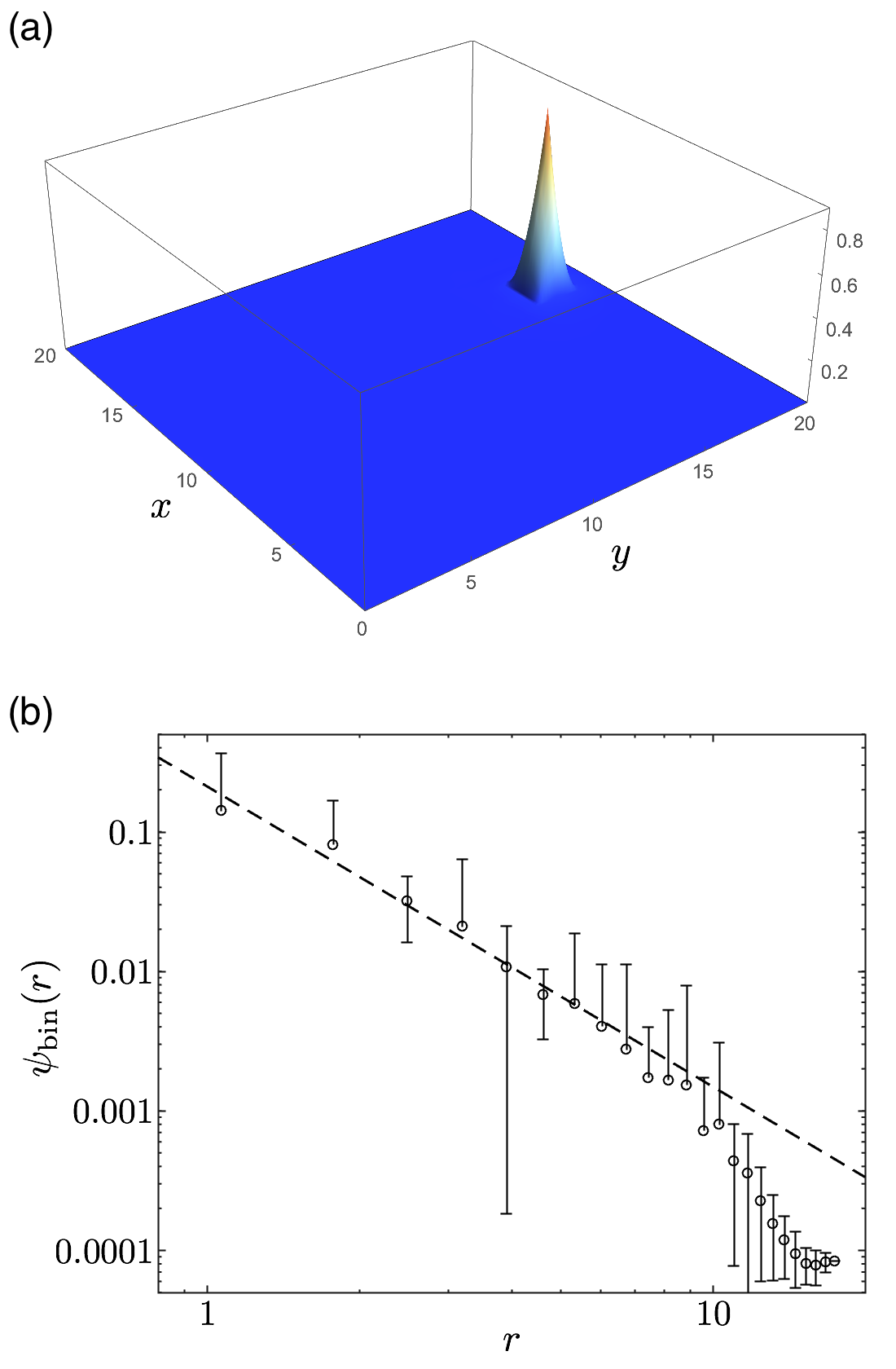}
	\caption{(a) The probability density projected to the $xy$ plane ($\sum_z|\psi(x,y,z)|^2$) of a rare state wave function with one flux quantum in the single-node Weyl model [here we are showing the rare state displayed in the dispersion in Fig.~\ref{fig:RS}(b)]. The rare state is chosen to be the eigenstate shown in Fig.~\ref{fig:RS}(b), $\phi = \pi$ (the state with low $\xi_L$ and $E/t \approx 0.06$). We can see the probability density is (quasi)localized about a rare region of the random potential. 
	(b) The decay of this rare state wavefunction. To see the decay behavior, we made a ``binned'' wave function with equally spaced bins (in distance $r$ from its maximum), and assigned the average value of the wave functions in the bin. The dashed line is a linear fit from data $r<6.4$ and has a slope of $-2.15$. This shows the power-law decay of the rare state wave function $\psi \sim 1/r^{-2.15}$.
    }
	\label{fig:RSWF}
\end{figure}
In this appendix, we quantitatively explore the dependence of the magnetic field on the quasilocalized rare wave functions. In Fig.~\ref{fig:RSWF}, we show a rare eigenstate for the single-node Weyl model with one magnetic flux. We find for short distances the power-law decay of the wave function is unaffected from the absence of a magnetic field and we find $\psi(r) \sim 1/r^{2.15}$ for $r \ll L$, which is in good agreement with the analytic prediction without a magnetic field $\psi(r) \sim 1/r^{2}$~\cite{Nandkishore14}. At larger $r$ we find that the wave function falls off faster than power law, here we attribute this to the single-node approximation we have made in the model, as this feature is not present at the same magnetic field strength in the two-node Weyl model.

\begin{figure}
\centering
	\vspace*{1mm}
	\includegraphics*[width=90mm,angle=0]{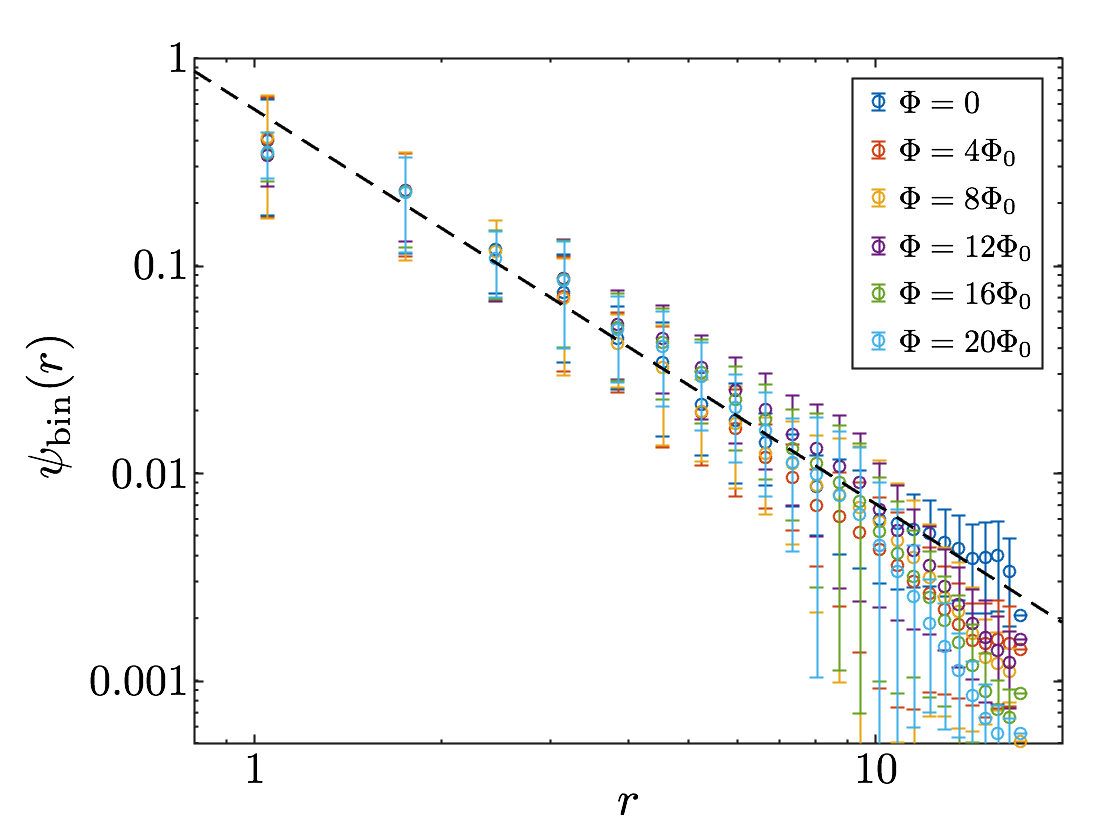}
	\caption{The decay of the rare state wavefunction in the presence of a magnetic field. Here, the rare state is found in the two Weyl model and is the same state we show in the dispersion in Fig.~\ref{fig:RS2}, $\phi = \pi$. The dashed line is the linear fit for the $\Phi = 0$ data, showing the power law decay $\psi \sim 1/r^{-1.9}$. The data shows even in the presence of a strong magnetic field (note that $\Phi = 20\Phi_0$ corresponds to $1/20$ flux per plaquette and the magnetic length $l_B/a \approx 1.78$) the small-$r$ behavior is not drastically changed. 
    }
	\label{fig:RSinB}
\end{figure}
To study the dependence on the magnetic field we focus on the two-node Weyl model and find a rare state with no magnetic field that decays like $\psi(r)\sim 1/r^{1.9}$ and then systematically increase the magnetic flux (focusing on the same rare state). We show the decay of the rare wave function in Fig.~\ref{fig:RSinB}; interestingly we find that the power-law decay of $\psi(r)$ at small $r \ll L$ is unaffected over a broad range of magnetic fields, whereas, the large-$r$ behavior falls off the power-law form more strongly for increasing magnetic flux. Nonetheless, even for the largest field strengths shown in Fig.~\ref{fig:RSinB} (with a magnetic length $l_B \approx 2 a$) the small-$r$ behavior is not dramatically affected. Thus, our results demonstrate the robustness of rare states in disordered Weyl semimetals to magnetic fields.

\begin{widetext}

\section{Short review on NL$\sigma$M}
\label{sec:AB}

In this appendix, we give a brief sketch on the derivation of the NL$\sigma$M [Eq.~\eqref{eq:F0}].
This is not a thorough derivation, nor an original work of the paper; rather, we summarize the concepts which are essential in understanding the main text. 
We use the supersymmetry method \cite{Efetov82,Efetov83,AltshulerPRL93,AltshulerPRB93,Efetov96} to evaluate the DOS autocorrelator. 
There are many literatures \cite{Efetov82,Efetov83,AltshulerPRL93,AltshulerPRB93,Efetov96} which worked on the details of this procedure and the symmetry of the NL$\sigma$M. Here, we closely follow especially Ref.~\cite{AltshulerPRB93}.

Let us state again the autocorrelator of the DOS, Eq.~\eqref{eq:K}: 
\begin{align}
	K(\omega, \delta \phi) &= \langle \nu(E, \phi) \, \nu (E+\omega, \phi+\delta \phi) \rangle \nonumber \\
	&= \frac{1}{V^2} \left\langle \sum_i \delta(E - E_i( \phi)) \, \sum_j \delta(E+\omega -E_j(\phi+\delta \phi)) \right\rangle .
	\label{eq:Kdef}
\end{align}
We denote the average $\langle \cdots \rangle$ over a range of energy, twist $0 \leq \phi < 2\pi$, and disorder realizations. 
We first express $K(\omega, \delta \phi)$ in terms of Green functions. 
When $\varphi_i ({\bf r})$ is the eigenstate of the Hamiltonian \eqref{eq:Hcont} with eigenvalue $E_i$, i.e., $H \varphi_i = E_i \varphi_i$, we can write the Green function as:
\begin{align}
	G^{R,A}_{E,\phi} ({\bf r}, {\bf r}') = \sum_i \frac{\varphi_i ({\bf r}) \varphi^*_i ({\bf r}')}{E - E_i(\phi) \pm i \delta}.
\end{align}
Now we can rewrite Eq.~(\ref{eq:Kdef}) as follows:
\begin{align}
	K(\omega, \delta \phi) &= -\frac{\Delta^2 \nu^2}{4 \pi^2} \int d{\bf r}\, d{\bf r}'
	\left\langle \left( G^A_{E,\phi} ({\bf r},{\bf r}) - G^R_{E,\phi} ({\bf r},{\bf r}) \right) 
	\left( G^A_{E+\omega,\phi+\delta \phi} ({\bf r}',{\bf r}') - G^R_{E+\omega,\phi+\delta \phi} ({\bf r}',{\bf r}') \right) \right\rangle.
	\label{eq:KGreen}
\end{align}
Here, we also substituted the inverse volume by the product of average level spacing ($\Delta$) and average DOS ($\nu$). 
Of the four terms in Eq.~(\ref{eq:KGreen}), $\langle G^A G^A \rangle$ and $\langle G^R G^R \rangle$ can be calculated from conventional perturbation theory. 
On the other hand, $\langle G^A G^R \rangle$ and $\langle G^R G^A \rangle$ are more difficult to handle perturbatively and we express these as a path integral over the eight-component supervector $\psi$:
\begin{align}
	G^A_{E,\phi} ({\bf r},{\bf r})  G^R_{E+\omega, \phi+\delta\phi} ({\bf r}',{\bf r}') = 
	\int \mathcal{D}\psi \mathcal{D} \bar\psi \,\, \psi^1_\alpha ({\bf r}) \bar\psi^1_\alpha({\bf r}) \, \psi^2_\beta ({\bf r}') \bar\psi^2_\beta({\bf r}')\, \textrm{exp}[-\mathcal L],
\end{align}
with the Lagrangian defined as:
\begin{align}
	\mathcal L = i \int d{\bf r} \, \bar\psi ({\bf r}) \left[
	- v \left( {\bf k} - \frac{e}{c} \tau_3 \,{\bf a} (\phi)\right)\cdot {\bf \sigma}  
	- V({\bf r}) + E + \frac{\omega}{2} - \frac{ (\omega + i \delta)}{2}\Lambda
	\right] \psi ({\bf r}).
\end{align}
$\psi$ is given as $\psi^T = \frac{1}{\sqrt 2}  \left( \chi^{1*}, \chi^{1}, S^{1*}, S^{1}, \chi^{2*}, \chi^{2}, S^{2*}, S^{2} \right)$ where the superscripts 1 and 2 are the advanced/retarded space; $\chi$'s are Grassmann variables and $S$'s are commuting variables. 
$\alpha$, $\beta$ are arbitrary components within the advanced and retarded spaces, respectively. 
$\Lambda$ is an $8 \times 8$ supermatrix defined as $\Lambda = \textrm{diag} (\mathbbm{1}_4 , -\mathbbm{1}_4)$ in this basis, and $\tau_3$ is the third Pauli matrix in $(\chi^*, \chi)$ and $(S^*, S)$ space.

Now, we can disorder average the theory exactly and replace $V \bar \psi \psi$ by $\frac{i}{4\pi \nu \tau} (\bar \psi \psi)^2$, where $\tau$ is the quasiparticle lifetime.
As the disorder-averaged Lagrangian now looks like an interacting theory without disorder, we can perform Hubbard-Stratonovich transformation introducing an $8 \times 8$ supermatrix field $Q({\bf r})$. 
Integrating out the original supervector field $\psi({\bf r})$, which is now quadratic, we obtain an effective action $F[Q]$:
\begin{align}
	F[Q] = \int d{\bf r} \left[ -\frac{1}{2} \str \ln \left(  - i {\mathcal H}_0  - \frac{i}{2} (\omega + i \delta) \Lambda + \frac{Q({\bf r})}{2 \tau} \right) + \frac{\pi \nu} {8 \tau} \str \, Q({\bf r})^2 \right],
	\label{eq:eff}
\end{align}
where $\mathcal H_0$ is:
\begin{align}
	\mathcal H_0 = v \left( {\bf k} - \frac{e}{c} \tau_3 \,{\bf a} (\phi)\right)\cdot {\bf \sigma} - E - \frac{\omega}{2} .
	\label{eq:H0}
\end{align}
Now, considering the small fluctuations around the saddle-point solution of Eq.~\eqref{eq:eff}, we obtain the desired NL$\sigma$M Eq.~\eqref{eq:F0}. 
Note that we have not used any details of the Hamiltonian up to this point, and emphasize this NL$\sigma$M is a general result. 

%%%

\section{Derivation of Chern-Simons term}
\label{sec:AC}

In the following, we show that when we assume a \emph{topological} Hamiltonian as in Eq.~(\ref{eq:Hcont}), the action [Eq.~(\ref{eq:F0})] has an additional Chern-Simons term which will significantly affect the value of the average level velocity. 
The strategy is to use the derivative expansion to Eq.~(\ref{eq:eff}) as in Ref.~\cite{Altland16}, where a similar term is derived in the replica framework. We stress that such a gradient expansion cannot capture the rare-region effects and we are focusing on deep in the diffusive regime.

First, we perform a similarity transform to change $F[Q]$ into $F[A]$. 
Here, we consider the $\omega = 0$ sector.
The long-wavelength expansion of Eq.~(\ref{eq:eff}) consists of a NL$\sigma$M coming from the large degeneracy of $\omega = 0$ sector, and $\omega$ linear terms. 
The Chern-Simons term we are interested in appears in the former and we can safely restrict ourselves to $\omega = 0$.
Performing the similarity transformation with $V$, and considering $Q({\bf r}) = V({\bf r}) \Lambda \bar{V}({\bf r})$,
\begin{align}
	F[A] &= \int d{\bf r} \left[ -\frac{1}{2} \str \ln \bar{V} \left(  - i \mathcal H_0   + \frac{Q({\bf r})}{2 \tau} \right) V+ \frac{\pi \nu} {8 \tau} \str \bar{V} Q V \bar{V} Q V \right] \nonumber \\
	&= \int d{\bf r} \left[ -\frac{1}{2} \str \ln \left( -i (\slashed{k}- i \slashed{A} - \epsilon) + \frac{\Lambda}{2 \tau} \right) \right].
\end{align}
The gauge-independent vector potential is $A_i = \bar{V} (\partial_i - i \tau_3 a_i) V$, and the slash notation is defined as $ \slashed{k} \equiv {\bf {k}} \cdot  \bm{\sigma}$.

Now, we are in a good position to expand the logarithm and write $F[A]$ as a power series of $A$. 
Only the action in the second power of $A$ is important in the derivation of $AdA$ Chern-Simons term:
\begin{align}
	F^{(2)}[A]= \int d{\bf r} \left[ -\frac{1}{4} \str (\epsilon - \slashed{k} - \frac{i \Lambda}{2 \tau})^{-1} \slashed{A}({\bf r}) (\epsilon - \slashed{k} - \frac{i \Lambda}{2 \tau})^{-1} \slashed{A}({\bf r}) \right].
\end{align}
We define $G_k = (\epsilon - \slashed{k} - \frac{i \Lambda}{2 \tau})^{-1} = \sum_{s = \pm} G_k^s P^s$, where $G_k^\pm = (\epsilon - \slashed{k} \mp \frac{i }{2 \tau})^{-1} $ and $P^\pm = (1 \pm \Lambda)/2$ the projection operator. 
Using the Moyal product expansion:
\begin{align}
	F^{(2)}[A] &\simeq -\frac{1}{4} \int d{\bf r} d{\bf k} \, \str \left(G_k \slashed{A}({\bf r}) - \frac{i}{2} \partial_{k_i} G_k \partial_{x_i} \slashed{A}({\bf r}) \right)  \left(G_k \slashed{A}({\bf r}) - \frac{i}{2} \partial_{k_i} G_k \partial_{x_i} \slashed{A}({\bf r})\right) \nonumber \\
	&= \cdots + \frac{i}{4} \int d{\bf r} d{\bf k} \, \str  \left( \partial_{k_i} G_k \right) \left( \partial_{x_i} \slashed{A}({\bf r}) \right) G_k \slashed{A} ({\bf r}) + \cdots \nonumber \\
	&= \cdots + \frac{i}{4} \sum_{s,s'} \int d{\bf r} d{\bf k} \, \str P^s \left( \partial_{k_i} G^s_k \right) \left( \partial_{x_i} A_j \right) \sigma_j P^{s'} G^{s'}_k A_m \sigma_m \nonumber \\
	&= \cdots + \frac{i}{4} \sum_{s,s'} \int d{\bf k} \,\tr \left( G^s_k \sigma_i G^s_k \sigma_j G^{s'}_k \sigma_m \right) \int d{\bf r} \, \str \left( P^s \partial_{x_i} A_j P^{s'} A_m \right)+ \cdots 
\end{align}
The leading order ``$\cdots$'' will become the $(\partial Q)^2$ term, and we focus on the subleading term written above. Note that we have used $\partial_{k_i} G^s_k = G^s_k \sigma_i G^s_k$ in the last equality.

We expand the $\textrm{tr}\, (G \sigma G \sigma G \sigma)$ term to obtain the desired Chern-Simons term:
\begin{align}
	F_{\textrm{CS}} &= -\frac{1}{2} \sum_{s,s'} F_{s s'} \int d{\bf r} \,\epsilon_{ijk} \,\str \left( P^s (\partial_{x_i} A_j) P^{s'} A_k \right), \nonumber \\
	F_{ss'} &= \int d{\bf k} \left(\epsilon_s^2 - k^2 \right)^{-2}
	\left(\epsilon_{s'}^2 - k^2 \right)^{-1}
	\left( \epsilon_s^2 \epsilon_{s'} - \frac{k^2}{3} (2\epsilon_s + \epsilon_{s'}) \right).
\end{align}
Here, $\epsilon_s \equiv \epsilon - \frac{i s}{2\tau}$.
The $F_{ss'}$ integral can be done directly as they nicely converge:
\begin{align}
	F_{ss'} = \frac{1}{8\pi} \left\{	\begin{array}{lcc}
	- is & & (s = s')\\
	4\epsilon \tau /3 & & (s \neq s')\\
	\end{array} \right. .
\end{align}
The $s = s'$ combination results in the action proportional to $i \epsilon_{ijk} \sum_{s = \pm} s \int d{\bf r} \, \str (A_i P^s \partial_{x_j} A_k P^s)$. 
Writing down this term explicitly,
\begin{align}
	F_{\textrm{CS}} = \frac{i}{16\pi} \sum_s s \int d{\bf r} \,\epsilon_{ijk} \,\str \left( P^s (\partial_{x_i} A_j) P^s A_k \right).
\end{align} 
We arrived at the form of Chern-Simons term as in Eq.~\eqref{eq:CS}.

Now we use the zero-mode approximation and ignore the spatial dependence of $V$, which gives $A_i = \bar{V} (-i \tau_3 a_i)V$, and use ${\bf a}$ as described in the main text ($\nabla \times {\bf a} = B \hat{z}$ with a curl-free term $(\Lambda \delta \phi/ L_z \hat{z})$).
\begin{align}
    F_{\textrm{CS}} &= -\frac{i}{16\pi} \sum_s s \int d{\bf r}  \,\str \left( P^s (\bar{V} B \tau_3 V) P^s (\bar{V} \frac{\Lambda \delta \phi}{L_z} \tau_3 V) \right) \nonumber \\
    &= -\frac{i B \delta \phi}{64\pi L_z} \sum_s s \int d{\bf r}  \,\str \left( (1+s \Lambda) (\bar{V} \tau_3 V) (1+s \Lambda)(\bar{V} \Lambda \tau_3 V) \right) \nonumber \\
    &= -\frac{i B \delta \phi}{32\pi L_z} \int d{\bf r}  \,\str \left( \Lambda (\bar{V} \tau_3 V) (\bar{V} \Lambda \tau_3 V) + (\bar{V} \tau_3 V) \Lambda (\bar{V} \Lambda \tau_3 V) \right) \nonumber \\
    &= -\frac{i B \delta \phi}{16\pi L_z} \int d{\bf r}  \,\str \left( \Lambda Q\right)
\end{align}
In the zero-mode approximation, the Chern-Simons term becomes proportional to $\str \left( \Lambda Q\right)$, which is the last term in Eq.~(\ref{eq:FQ}).

\end{widetext}

\bibliographystyle{apsrev4-1}
\bibliography{Weyl}

%merlin.mbs apsrev4-1.bst 2010-07-25 4.21a (PWD, AO, DPC) hacked
%Control: key (0)
%Control: author (72) initials jnrlst
%Control: editor formatted (1) identically to author
%Control: production of article title (-1) disabled
%Control: page (0) single
%Control: year (1) truncated
%Control: production of eprint (0) enabled
\begin{thebibliography}{76}%
\makeatletter
\providecommand \@ifxundefined [1]{%
 \@ifx{#1\undefined}
}%
\providecommand \@ifnum [1]{%
 \ifnum #1\expandafter \@firstoftwo
 \else \expandafter \@secondoftwo
 \fi
}%
\providecommand \@ifx [1]{%
 \ifx #1\expandafter \@firstoftwo
 \else \expandafter \@secondoftwo
 \fi
}%
\providecommand \natexlab [1]{#1}%
\providecommand \enquote  [1]{``#1''}%
\providecommand \bibnamefont  [1]{#1}%
\providecommand \bibfnamefont [1]{#1}%
\providecommand \citenamefont [1]{#1}%
\providecommand \href@noop [0]{\@secondoftwo}%
\providecommand \href [0]{\begingroup \@sanitize@url \@href}%
\providecommand \@href[1]{\@@startlink{#1}\@@href}%
\providecommand \@@href[1]{\endgroup#1\@@endlink}%
\providecommand \@sanitize@url [0]{\catcode `\\12\catcode `\$12\catcode
  `\&12\catcode `\#12\catcode `\^12\catcode `\_12\catcode `\%12\relax}%
\providecommand \@@startlink[1]{}%
\providecommand \@@endlink[0]{}%
\providecommand \url  [0]{\begingroup\@sanitize@url \@url }%
\providecommand \@url [1]{\endgroup\@href {#1}{\urlprefix }}%
\providecommand \urlprefix  [0]{URL }%
\providecommand \Eprint [0]{\href }%
\providecommand \doibase [0]{http://dx.doi.org/}%
\providecommand \selectlanguage [0]{\@gobble}%
\providecommand \bibinfo  [0]{\@secondoftwo}%
\providecommand \bibfield  [0]{\@secondoftwo}%
\providecommand \translation [1]{[#1]}%
\providecommand \BibitemOpen [0]{}%
\providecommand \bibitemStop [0]{}%
\providecommand \bibitemNoStop [0]{.\EOS\space}%
\providecommand \EOS [0]{\spacefactor3000\relax}%
\providecommand \BibitemShut  [1]{\csname bibitem#1\endcsname}%
\let\auto@bib@innerbib\@empty
%</preamble>
\bibitem [{\citenamefont {Adler}(1969)}]{Adler69}%
  \BibitemOpen
  \bibfield  {author} {\bibinfo {author} {\bibfnamefont {S.~L.}\ \bibnamefont
  {Adler}},\ }\href {\doibase 10.1103/PhysRev.177.2426} {\bibfield  {journal}
  {\bibinfo  {journal} {Phys. Rev.}\ }\textbf {\bibinfo {volume} {177}},\
  \bibinfo {pages} {2426} (\bibinfo {year} {1969})}\BibitemShut {NoStop}%
\bibitem [{\citenamefont {Bell}\ and\ \citenamefont {Jackiw}(1969)}]{Bell69}%
  \BibitemOpen
  \bibfield  {author} {\bibinfo {author} {\bibfnamefont {J.~S.}\ \bibnamefont
  {Bell}}\ and\ \bibinfo {author} {\bibfnamefont {R.}~\bibnamefont {Jackiw}},\
  }\href {\doibase 10.1007/BF02823296} {\bibfield  {journal} {\bibinfo
  {journal} {Il Nuovo Cimento A (1965-1970)}\ }\textbf {\bibinfo {volume}
  {60}},\ \bibinfo {pages} {47} (\bibinfo {year} {1969})}\BibitemShut {NoStop}%
\bibitem [{\citenamefont {Liu}\ \emph {et~al.}(2014{\natexlab{a}})\citenamefont
  {Liu}, \citenamefont {Zhou}, \citenamefont {Zhang}, \citenamefont {Wang},
  \citenamefont {Weng}, \citenamefont {Prabhakaran}, \citenamefont {Mo},
  \citenamefont {Shen}, \citenamefont {Fang}, \citenamefont {Dai},
  \citenamefont {Hussain},\ and\ \citenamefont {Chen}}]{Liu14Science}%
  \BibitemOpen
  \bibfield  {author} {\bibinfo {author} {\bibfnamefont {Z.~K.}\ \bibnamefont
  {Liu}}, \bibinfo {author} {\bibfnamefont {B.}~\bibnamefont {Zhou}}, \bibinfo
  {author} {\bibfnamefont {Y.}~\bibnamefont {Zhang}}, \bibinfo {author}
  {\bibfnamefont {Z.~J.}\ \bibnamefont {Wang}}, \bibinfo {author}
  {\bibfnamefont {H.~M.}\ \bibnamefont {Weng}}, \bibinfo {author}
  {\bibfnamefont {D.}~\bibnamefont {Prabhakaran}}, \bibinfo {author}
  {\bibfnamefont {S.-K.}\ \bibnamefont {Mo}}, \bibinfo {author} {\bibfnamefont
  {Z.~X.}\ \bibnamefont {Shen}}, \bibinfo {author} {\bibfnamefont
  {Z.}~\bibnamefont {Fang}}, \bibinfo {author} {\bibfnamefont {X.}~\bibnamefont
  {Dai}}, \bibinfo {author} {\bibfnamefont {Z.}~\bibnamefont {Hussain}}, \ and\
  \bibinfo {author} {\bibfnamefont {Y.~L.}\ \bibnamefont {Chen}},\ }\href
  {\doibase 10.1126/science.1245085} {\bibfield  {journal} {\bibinfo  {journal}
  {Science}\ }\textbf {\bibinfo {volume} {343}},\ \bibinfo {pages} {864}
  (\bibinfo {year} {2014}{\natexlab{a}})},\ \Eprint
  {http://arxiv.org/abs/http://science.sciencemag.org/content/343/6173/864.full.pdf}
  {http://science.sciencemag.org/content/343/6173/864.full.pdf} \BibitemShut
  {NoStop}%
\bibitem [{\citenamefont {Xu}\ \emph {et~al.}(2015{\natexlab{a}})\citenamefont
  {Xu}, \citenamefont {Liu}, \citenamefont {Kushwaha}, \citenamefont {Sankar},
  \citenamefont {Krizan}, \citenamefont {Belopolski}, \citenamefont {Neupane},
  \citenamefont {Bian}, \citenamefont {Alidoust}, \citenamefont {Chang},
  \citenamefont {Jeng}, \citenamefont {Huang}, \citenamefont {Tsai},
  \citenamefont {Lin}, \citenamefont {Shibayev}, \citenamefont {Chou},
  \citenamefont {Cava},\ and\ \citenamefont {Hasan}}]{Xu15Science2}%
  \BibitemOpen
  \bibfield  {author} {\bibinfo {author} {\bibfnamefont {S.-Y.}\ \bibnamefont
  {Xu}}, \bibinfo {author} {\bibfnamefont {C.}~\bibnamefont {Liu}}, \bibinfo
  {author} {\bibfnamefont {S.~K.}\ \bibnamefont {Kushwaha}}, \bibinfo {author}
  {\bibfnamefont {R.}~\bibnamefont {Sankar}}, \bibinfo {author} {\bibfnamefont
  {J.~W.}\ \bibnamefont {Krizan}}, \bibinfo {author} {\bibfnamefont
  {I.}~\bibnamefont {Belopolski}}, \bibinfo {author} {\bibfnamefont
  {M.}~\bibnamefont {Neupane}}, \bibinfo {author} {\bibfnamefont
  {G.}~\bibnamefont {Bian}}, \bibinfo {author} {\bibfnamefont {N.}~\bibnamefont
  {Alidoust}}, \bibinfo {author} {\bibfnamefont {T.-R.}\ \bibnamefont {Chang}},
  \bibinfo {author} {\bibfnamefont {H.-T.}\ \bibnamefont {Jeng}}, \bibinfo
  {author} {\bibfnamefont {C.-Y.}\ \bibnamefont {Huang}}, \bibinfo {author}
  {\bibfnamefont {W.-F.}\ \bibnamefont {Tsai}}, \bibinfo {author}
  {\bibfnamefont {H.}~\bibnamefont {Lin}}, \bibinfo {author} {\bibfnamefont
  {P.~P.}\ \bibnamefont {Shibayev}}, \bibinfo {author} {\bibfnamefont {F.-C.}\
  \bibnamefont {Chou}}, \bibinfo {author} {\bibfnamefont {R.~J.}\ \bibnamefont
  {Cava}}, \ and\ \bibinfo {author} {\bibfnamefont {M.~Z.}\ \bibnamefont
  {Hasan}},\ }\href {\doibase 10.1126/science.1256742} {\bibfield  {journal}
  {\bibinfo  {journal} {Science}\ }\textbf {\bibinfo {volume} {347}},\ \bibinfo
  {pages} {294} (\bibinfo {year} {2015}{\natexlab{a}})},\ \Eprint
  {http://arxiv.org/abs/http://science.sciencemag.org/content/347/6219/294.full.pdf}
  {http://science.sciencemag.org/content/347/6219/294.full.pdf} \BibitemShut
  {NoStop}%
\bibitem [{\citenamefont {Borisenko}\ \emph {et~al.}(2014)\citenamefont
  {Borisenko}, \citenamefont {Gibson}, \citenamefont {Evtushinsky},
  \citenamefont {Zabolotnyy}, \citenamefont {B\"uchner},\ and\ \citenamefont
  {Cava}}]{Borisenko14}%
  \BibitemOpen
  \bibfield  {author} {\bibinfo {author} {\bibfnamefont {S.}~\bibnamefont
  {Borisenko}}, \bibinfo {author} {\bibfnamefont {Q.}~\bibnamefont {Gibson}},
  \bibinfo {author} {\bibfnamefont {D.}~\bibnamefont {Evtushinsky}}, \bibinfo
  {author} {\bibfnamefont {V.}~\bibnamefont {Zabolotnyy}}, \bibinfo {author}
  {\bibfnamefont {B.}~\bibnamefont {B\"uchner}}, \ and\ \bibinfo {author}
  {\bibfnamefont {R.~J.}\ \bibnamefont {Cava}},\ }\href {\doibase
  10.1103/PhysRevLett.113.027603} {\bibfield  {journal} {\bibinfo  {journal}
  {Phys. Rev. Lett.}\ }\textbf {\bibinfo {volume} {113}},\ \bibinfo {pages}
  {027603} (\bibinfo {year} {2014})}\BibitemShut {NoStop}%
\bibitem [{\citenamefont {Liu}\ \emph {et~al.}(2014{\natexlab{b}})\citenamefont
  {Liu}, \citenamefont {Jiang}, \citenamefont {Zhou}, \citenamefont {Wang},
  \citenamefont {Zhang}, \citenamefont {Weng}, \citenamefont {Prabhakaran},
  \citenamefont {Mo}, \citenamefont {Peng}, \citenamefont {Dudin},
  \citenamefont {Kim}, \citenamefont {Hoesch}, \citenamefont {Fang},
  \citenamefont {Dai}, \citenamefont {Shen}, \citenamefont {Feng},
  \citenamefont {Hussain},\ and\ \citenamefont {Chen}}]{Liu14NatMat}%
  \BibitemOpen
  \bibfield  {author} {\bibinfo {author} {\bibfnamefont {Z.~K.}\ \bibnamefont
  {Liu}}, \bibinfo {author} {\bibfnamefont {J.}~\bibnamefont {Jiang}}, \bibinfo
  {author} {\bibfnamefont {B.}~\bibnamefont {Zhou}}, \bibinfo {author}
  {\bibfnamefont {Z.~J.}\ \bibnamefont {Wang}}, \bibinfo {author}
  {\bibfnamefont {Y.}~\bibnamefont {Zhang}}, \bibinfo {author} {\bibfnamefont
  {H.~M.}\ \bibnamefont {Weng}}, \bibinfo {author} {\bibfnamefont
  {D.}~\bibnamefont {Prabhakaran}}, \bibinfo {author} {\bibfnamefont {S.-K.}\
  \bibnamefont {Mo}}, \bibinfo {author} {\bibfnamefont {H.}~\bibnamefont
  {Peng}}, \bibinfo {author} {\bibfnamefont {P.}~\bibnamefont {Dudin}},
  \bibinfo {author} {\bibfnamefont {T.}~\bibnamefont {Kim}}, \bibinfo {author}
  {\bibfnamefont {M.}~\bibnamefont {Hoesch}}, \bibinfo {author} {\bibfnamefont
  {Z.}~\bibnamefont {Fang}}, \bibinfo {author} {\bibfnamefont {X.}~\bibnamefont
  {Dai}}, \bibinfo {author} {\bibfnamefont {Z.~X.}\ \bibnamefont {Shen}},
  \bibinfo {author} {\bibfnamefont {D.~L.}\ \bibnamefont {Feng}}, \bibinfo
  {author} {\bibfnamefont {Z.}~\bibnamefont {Hussain}}, \ and\ \bibinfo
  {author} {\bibfnamefont {Y.~L.}\ \bibnamefont {Chen}},\ }\href
  {http://dx.doi.org/10.1038/nmat3990} {\bibfield  {journal} {\bibinfo
  {journal} {Nature Materials}\ }\textbf {\bibinfo {volume} {13}},\ \bibinfo
  {pages} {677 EP } (\bibinfo {year} {2014}{\natexlab{b}})}\BibitemShut
  {NoStop}%
\bibitem [{\citenamefont {Neupane}\ \emph {et~al.}(2014)\citenamefont
  {Neupane}, \citenamefont {Xu}, \citenamefont {Sankar}, \citenamefont
  {Alidoust}, \citenamefont {Bian}, \citenamefont {Liu}, \citenamefont
  {Belopolski}, \citenamefont {Chang}, \citenamefont {Jeng}, \citenamefont
  {Lin}, \citenamefont {Bansil}, \citenamefont {Chou},\ and\ \citenamefont
  {Hasan}}]{Neupane14}%
  \BibitemOpen
  \bibfield  {author} {\bibinfo {author} {\bibfnamefont {M.}~\bibnamefont
  {Neupane}}, \bibinfo {author} {\bibfnamefont {S.-Y.}\ \bibnamefont {Xu}},
  \bibinfo {author} {\bibfnamefont {R.}~\bibnamefont {Sankar}}, \bibinfo
  {author} {\bibfnamefont {N.}~\bibnamefont {Alidoust}}, \bibinfo {author}
  {\bibfnamefont {G.}~\bibnamefont {Bian}}, \bibinfo {author} {\bibfnamefont
  {C.}~\bibnamefont {Liu}}, \bibinfo {author} {\bibfnamefont {I.}~\bibnamefont
  {Belopolski}}, \bibinfo {author} {\bibfnamefont {T.-R.}\ \bibnamefont
  {Chang}}, \bibinfo {author} {\bibfnamefont {H.-T.}\ \bibnamefont {Jeng}},
  \bibinfo {author} {\bibfnamefont {H.}~\bibnamefont {Lin}}, \bibinfo {author}
  {\bibfnamefont {A.}~\bibnamefont {Bansil}}, \bibinfo {author} {\bibfnamefont
  {F.}~\bibnamefont {Chou}}, \ and\ \bibinfo {author} {\bibfnamefont {M.~Z.}\
  \bibnamefont {Hasan}},\ }\href {http://dx.doi.org/10.1038/ncomms4786}
  {\bibfield  {journal} {\bibinfo  {journal} {Nature Communications}\ }\textbf
  {\bibinfo {volume} {5}},\ \bibinfo {pages} {3786 EP } (\bibinfo {year}
  {2014})}\BibitemShut {NoStop}%
\bibitem [{\citenamefont {Lv}\ \emph {et~al.}(2015{\natexlab{a}})\citenamefont
  {Lv}, \citenamefont {Weng}, \citenamefont {Fu}, \citenamefont {Wang},
  \citenamefont {Miao}, \citenamefont {Ma}, \citenamefont {Richard},
  \citenamefont {Huang}, \citenamefont {Zhao}, \citenamefont {Chen},
  \citenamefont {Fang}, \citenamefont {Dai}, \citenamefont {Qian},\ and\
  \citenamefont {Ding}}]{Lv15PRX}%
  \BibitemOpen
  \bibfield  {author} {\bibinfo {author} {\bibfnamefont {B.~Q.}\ \bibnamefont
  {Lv}}, \bibinfo {author} {\bibfnamefont {H.~M.}\ \bibnamefont {Weng}},
  \bibinfo {author} {\bibfnamefont {B.~B.}\ \bibnamefont {Fu}}, \bibinfo
  {author} {\bibfnamefont {X.~P.}\ \bibnamefont {Wang}}, \bibinfo {author}
  {\bibfnamefont {H.}~\bibnamefont {Miao}}, \bibinfo {author} {\bibfnamefont
  {J.}~\bibnamefont {Ma}}, \bibinfo {author} {\bibfnamefont {P.}~\bibnamefont
  {Richard}}, \bibinfo {author} {\bibfnamefont {X.~C.}\ \bibnamefont {Huang}},
  \bibinfo {author} {\bibfnamefont {L.~X.}\ \bibnamefont {Zhao}}, \bibinfo
  {author} {\bibfnamefont {G.~F.}\ \bibnamefont {Chen}}, \bibinfo {author}
  {\bibfnamefont {Z.}~\bibnamefont {Fang}}, \bibinfo {author} {\bibfnamefont
  {X.}~\bibnamefont {Dai}}, \bibinfo {author} {\bibfnamefont {T.}~\bibnamefont
  {Qian}}, \ and\ \bibinfo {author} {\bibfnamefont {H.}~\bibnamefont {Ding}},\
  }\href {\doibase 10.1103/PhysRevX.5.031013} {\bibfield  {journal} {\bibinfo
  {journal} {Phys. Rev. X}\ }\textbf {\bibinfo {volume} {5}},\ \bibinfo {pages}
  {031013} (\bibinfo {year} {2015}{\natexlab{a}})}\BibitemShut {NoStop}%
\bibitem [{\citenamefont {Lv}\ \emph {et~al.}(2015{\natexlab{b}})\citenamefont
  {Lv}, \citenamefont {Xu}, \citenamefont {Weng}, \citenamefont {Ma},
  \citenamefont {Richard}, \citenamefont {Huang}, \citenamefont {Zhao},
  \citenamefont {Chen}, \citenamefont {Matt}, \citenamefont {Bisti},
  \citenamefont {Strocov}, \citenamefont {Mesot}, \citenamefont {Fang},
  \citenamefont {Dai}, \citenamefont {Qian}, \citenamefont {Shi},\ and\
  \citenamefont {Ding}}]{Lv15NatPhys}%
  \BibitemOpen
  \bibfield  {author} {\bibinfo {author} {\bibfnamefont {B.~Q.}\ \bibnamefont
  {Lv}}, \bibinfo {author} {\bibfnamefont {N.}~\bibnamefont {Xu}}, \bibinfo
  {author} {\bibfnamefont {H.~M.}\ \bibnamefont {Weng}}, \bibinfo {author}
  {\bibfnamefont {J.~Z.}\ \bibnamefont {Ma}}, \bibinfo {author} {\bibfnamefont
  {P.}~\bibnamefont {Richard}}, \bibinfo {author} {\bibfnamefont {X.~C.}\
  \bibnamefont {Huang}}, \bibinfo {author} {\bibfnamefont {L.~X.}\ \bibnamefont
  {Zhao}}, \bibinfo {author} {\bibfnamefont {G.~F.}\ \bibnamefont {Chen}},
  \bibinfo {author} {\bibfnamefont {C.~E.}\ \bibnamefont {Matt}}, \bibinfo
  {author} {\bibfnamefont {F.}~\bibnamefont {Bisti}}, \bibinfo {author}
  {\bibfnamefont {V.~N.}\ \bibnamefont {Strocov}}, \bibinfo {author}
  {\bibfnamefont {J.}~\bibnamefont {Mesot}}, \bibinfo {author} {\bibfnamefont
  {Z.}~\bibnamefont {Fang}}, \bibinfo {author} {\bibfnamefont {X.}~\bibnamefont
  {Dai}}, \bibinfo {author} {\bibfnamefont {T.}~\bibnamefont {Qian}}, \bibinfo
  {author} {\bibfnamefont {M.}~\bibnamefont {Shi}}, \ and\ \bibinfo {author}
  {\bibfnamefont {H.}~\bibnamefont {Ding}},\ }\href
  {http://dx.doi.org/10.1038/nphys3426} {\bibfield  {journal} {\bibinfo
  {journal} {Nature Physics}\ }\textbf {\bibinfo {volume} {11}},\ \bibinfo
  {pages} {724 EP } (\bibinfo {year} {2015}{\natexlab{b}})}\BibitemShut
  {NoStop}%
\bibitem [{\citenamefont {Xu}\ \emph {et~al.}(2015{\natexlab{b}})\citenamefont
  {Xu}, \citenamefont {Belopolski}, \citenamefont {Alidoust}, \citenamefont
  {Neupane}, \citenamefont {Bian}, \citenamefont {Zhang}, \citenamefont
  {Sankar}, \citenamefont {Chang}, \citenamefont {Yuan}, \citenamefont {Lee},
  \citenamefont {Huang}, \citenamefont {Zheng}, \citenamefont {Ma},
  \citenamefont {Sanchez}, \citenamefont {Wang}, \citenamefont {Bansil},
  \citenamefont {Chou}, \citenamefont {Shibayev}, \citenamefont {Lin},
  \citenamefont {Jia},\ and\ \citenamefont {Hasan}}]{Xu15Science1}%
  \BibitemOpen
  \bibfield  {author} {\bibinfo {author} {\bibfnamefont {S.-Y.}\ \bibnamefont
  {Xu}}, \bibinfo {author} {\bibfnamefont {I.}~\bibnamefont {Belopolski}},
  \bibinfo {author} {\bibfnamefont {N.}~\bibnamefont {Alidoust}}, \bibinfo
  {author} {\bibfnamefont {M.}~\bibnamefont {Neupane}}, \bibinfo {author}
  {\bibfnamefont {G.}~\bibnamefont {Bian}}, \bibinfo {author} {\bibfnamefont
  {C.}~\bibnamefont {Zhang}}, \bibinfo {author} {\bibfnamefont
  {R.}~\bibnamefont {Sankar}}, \bibinfo {author} {\bibfnamefont
  {G.}~\bibnamefont {Chang}}, \bibinfo {author} {\bibfnamefont
  {Z.}~\bibnamefont {Yuan}}, \bibinfo {author} {\bibfnamefont {C.-C.}\
  \bibnamefont {Lee}}, \bibinfo {author} {\bibfnamefont {S.-M.}\ \bibnamefont
  {Huang}}, \bibinfo {author} {\bibfnamefont {H.}~\bibnamefont {Zheng}},
  \bibinfo {author} {\bibfnamefont {J.}~\bibnamefont {Ma}}, \bibinfo {author}
  {\bibfnamefont {D.~S.}\ \bibnamefont {Sanchez}}, \bibinfo {author}
  {\bibfnamefont {B.}~\bibnamefont {Wang}}, \bibinfo {author} {\bibfnamefont
  {A.}~\bibnamefont {Bansil}}, \bibinfo {author} {\bibfnamefont
  {F.}~\bibnamefont {Chou}}, \bibinfo {author} {\bibfnamefont {P.~P.}\
  \bibnamefont {Shibayev}}, \bibinfo {author} {\bibfnamefont {H.}~\bibnamefont
  {Lin}}, \bibinfo {author} {\bibfnamefont {S.}~\bibnamefont {Jia}}, \ and\
  \bibinfo {author} {\bibfnamefont {M.~Z.}\ \bibnamefont {Hasan}},\ }\href
  {\doibase 10.1126/science.aaa9297} {\bibfield  {journal} {\bibinfo  {journal}
  {Science}\ }\textbf {\bibinfo {volume} {349}},\ \bibinfo {pages} {613}
  (\bibinfo {year} {2015}{\natexlab{b}})},\ \Eprint
  {http://arxiv.org/abs/http://science.sciencemag.org/content/349/6248/613.full.pdf}
  {http://science.sciencemag.org/content/349/6248/613.full.pdf} \BibitemShut
  {NoStop}%
\bibitem [{\citenamefont {Yang}\ \emph
  {et~al.}(2015{\natexlab{a}})\citenamefont {Yang}, \citenamefont {Liu},
  \citenamefont {Sun}, \citenamefont {Peng}, \citenamefont {Yang},
  \citenamefont {Zhang}, \citenamefont {Zhou}, \citenamefont {Zhang},
  \citenamefont {Guo}, \citenamefont {Rahn}, \citenamefont {Prabhakaran},
  \citenamefont {Hussain}, \citenamefont {Mo}, \citenamefont {Felser},
  \citenamefont {Yan},\ and\ \citenamefont {Chen}}]{YangL15}%
  \BibitemOpen
  \bibfield  {author} {\bibinfo {author} {\bibfnamefont {L.~X.}\ \bibnamefont
  {Yang}}, \bibinfo {author} {\bibfnamefont {Z.~K.}\ \bibnamefont {Liu}},
  \bibinfo {author} {\bibfnamefont {Y.}~\bibnamefont {Sun}}, \bibinfo {author}
  {\bibfnamefont {H.}~\bibnamefont {Peng}}, \bibinfo {author} {\bibfnamefont
  {H.~F.}\ \bibnamefont {Yang}}, \bibinfo {author} {\bibfnamefont
  {T.}~\bibnamefont {Zhang}}, \bibinfo {author} {\bibfnamefont
  {B.}~\bibnamefont {Zhou}}, \bibinfo {author} {\bibfnamefont {Y.}~\bibnamefont
  {Zhang}}, \bibinfo {author} {\bibfnamefont {Y.~F.}\ \bibnamefont {Guo}},
  \bibinfo {author} {\bibfnamefont {M.}~\bibnamefont {Rahn}}, \bibinfo {author}
  {\bibfnamefont {D.}~\bibnamefont {Prabhakaran}}, \bibinfo {author}
  {\bibfnamefont {Z.}~\bibnamefont {Hussain}}, \bibinfo {author} {\bibfnamefont
  {S.~K.}\ \bibnamefont {Mo}}, \bibinfo {author} {\bibfnamefont
  {C.}~\bibnamefont {Felser}}, \bibinfo {author} {\bibfnamefont
  {B.}~\bibnamefont {Yan}}, \ and\ \bibinfo {author} {\bibfnamefont {Y.~L.}\
  \bibnamefont {Chen}},\ }\href {http://dx.doi.org/10.1038/nphys3425}
  {\bibfield  {journal} {\bibinfo  {journal} {Nature Physics}\ }\textbf
  {\bibinfo {volume} {11}},\ \bibinfo {pages} {728 EP } (\bibinfo {year}
  {2015}{\natexlab{a}})}\BibitemShut {NoStop}%
\bibitem [{\citenamefont {Xu}\ \emph {et~al.}(2015{\natexlab{c}})\citenamefont
  {Xu}, \citenamefont {Alidoust}, \citenamefont {Belopolski}, \citenamefont
  {Yuan}, \citenamefont {Bian}, \citenamefont {Chang}, \citenamefont {Zheng},
  \citenamefont {Strocov}, \citenamefont {Sanchez}, \citenamefont {Chang},
  \citenamefont {Zhang}, \citenamefont {Mou}, \citenamefont {Wu}, \citenamefont
  {Huang}, \citenamefont {Lee}, \citenamefont {Huang}, \citenamefont {Wang},
  \citenamefont {Bansil}, \citenamefont {Jeng}, \citenamefont {Neuport},
  \citenamefont {Kaminski}, \citenamefont {Lin},\ and\ \citenamefont
  {Hasan}}]{xu2015discovery}%
  \BibitemOpen
  \bibfield  {author} {\bibinfo {author} {\bibfnamefont {S.-Y.}\ \bibnamefont
  {Xu}}, \bibinfo {author} {\bibfnamefont {N.}~\bibnamefont {Alidoust}},
  \bibinfo {author} {\bibfnamefont {I.}~\bibnamefont {Belopolski}}, \bibinfo
  {author} {\bibfnamefont {Z.}~\bibnamefont {Yuan}}, \bibinfo {author}
  {\bibfnamefont {G.}~\bibnamefont {Bian}}, \bibinfo {author} {\bibfnamefont
  {T.-R.}\ \bibnamefont {Chang}}, \bibinfo {author} {\bibfnamefont
  {H.}~\bibnamefont {Zheng}}, \bibinfo {author} {\bibfnamefont {V.~N.}\
  \bibnamefont {Strocov}}, \bibinfo {author} {\bibfnamefont {D.~S.}\
  \bibnamefont {Sanchez}}, \bibinfo {author} {\bibfnamefont {G.}~\bibnamefont
  {Chang}}, \bibinfo {author} {\bibfnamefont {C.}~\bibnamefont {Zhang}},
  \bibinfo {author} {\bibfnamefont {D.}~\bibnamefont {Mou}}, \bibinfo {author}
  {\bibfnamefont {Y.}~\bibnamefont {Wu}}, \bibinfo {author} {\bibfnamefont
  {L.}~\bibnamefont {Huang}}, \bibinfo {author} {\bibfnamefont {C.-C.}\
  \bibnamefont {Lee}}, \bibinfo {author} {\bibfnamefont {S.-M.}\ \bibnamefont
  {Huang}}, \bibinfo {author} {\bibfnamefont {B.}~\bibnamefont {Wang}},
  \bibinfo {author} {\bibfnamefont {A.}~\bibnamefont {Bansil}}, \bibinfo
  {author} {\bibfnamefont {H.-T.}\ \bibnamefont {Jeng}}, \bibinfo {author}
  {\bibfnamefont {T.}~\bibnamefont {Neuport}}, \bibinfo {author} {\bibfnamefont
  {A.}~\bibnamefont {Kaminski}}, \bibinfo {author} {\bibfnamefont
  {H.}~\bibnamefont {Lin}}, \ and\ \bibinfo {author} {\bibfnamefont {M.~Z.}\
  \bibnamefont {Hasan}},\ }\href {https://www.nature.com/articles/nphys3437}
  {\bibfield  {journal} {\bibinfo  {journal} {Nature Physics}\ }\textbf
  {\bibinfo {volume} {11}},\ \bibinfo {pages} {748} (\bibinfo {year}
  {2015}{\natexlab{c}})}\BibitemShut {NoStop}%
\bibitem [{\citenamefont {Xu}\ \emph {et~al.}(2016)\citenamefont {Xu},
  \citenamefont {Weng}, \citenamefont {Lv}, \citenamefont {Matt}, \citenamefont
  {Park}, \citenamefont {Bisti}, \citenamefont {Strocov}, \citenamefont
  {Gawryluk}, \citenamefont {Pomjakushina}, \citenamefont {Conder},
  \citenamefont {Plumb}, \citenamefont {Radovic}, \citenamefont {Aut\`{e}s},
  \citenamefont {Yazyev}, \citenamefont {Fang}, \citenamefont {Dai},
  \citenamefont {Qian}, \citenamefont {Mesot}, \citenamefont {Ding},\ and\
  \citenamefont {Shi}}]{xu2016observation}%
  \BibitemOpen
  \bibfield  {author} {\bibinfo {author} {\bibfnamefont {N.}~\bibnamefont
  {Xu}}, \bibinfo {author} {\bibfnamefont {H.~M.}\ \bibnamefont {Weng}},
  \bibinfo {author} {\bibfnamefont {B.~Q.}\ \bibnamefont {Lv}}, \bibinfo
  {author} {\bibfnamefont {C.~E.}\ \bibnamefont {Matt}}, \bibinfo {author}
  {\bibfnamefont {J.}~\bibnamefont {Park}}, \bibinfo {author} {\bibfnamefont
  {F.}~\bibnamefont {Bisti}}, \bibinfo {author} {\bibfnamefont {V.~N.}\
  \bibnamefont {Strocov}}, \bibinfo {author} {\bibfnamefont {D.}~\bibnamefont
  {Gawryluk}}, \bibinfo {author} {\bibfnamefont {E.}~\bibnamefont
  {Pomjakushina}}, \bibinfo {author} {\bibfnamefont {K.}~\bibnamefont
  {Conder}}, \bibinfo {author} {\bibfnamefont {N.~C.}\ \bibnamefont {Plumb}},
  \bibinfo {author} {\bibfnamefont {M.}~\bibnamefont {Radovic}}, \bibinfo
  {author} {\bibfnamefont {G.}~\bibnamefont {Aut\`{e}s}}, \bibinfo {author}
  {\bibfnamefont {O.~V.}\ \bibnamefont {Yazyev}}, \bibinfo {author}
  {\bibfnamefont {Z.}~\bibnamefont {Fang}}, \bibinfo {author} {\bibfnamefont
  {X.}~\bibnamefont {Dai}}, \bibinfo {author} {\bibfnamefont {T.}~\bibnamefont
  {Qian}}, \bibinfo {author} {\bibfnamefont {J.}~\bibnamefont {Mesot}},
  \bibinfo {author} {\bibfnamefont {H.}~\bibnamefont {Ding}}, \ and\ \bibinfo
  {author} {\bibfnamefont {M.}~\bibnamefont {Shi}},\ }\href@noop {} {\bibfield
  {journal} {\bibinfo  {journal} {Nature Communications}\ }\textbf {\bibinfo
  {volume} {7}},\ \bibinfo {pages} {11006} (\bibinfo {year}
  {2016})}\BibitemShut {NoStop}%
\bibitem [{\citenamefont {Nielsen}\ and\ \citenamefont
  {Ninomiya}(1981{\natexlab{a}})}]{Nielsen81NPB1}%
  \BibitemOpen
  \bibfield  {author} {\bibinfo {author} {\bibfnamefont {H.}~\bibnamefont
  {Nielsen}}\ and\ \bibinfo {author} {\bibfnamefont {M.}~\bibnamefont
  {Ninomiya}},\ }\href {\doibase https://doi.org/10.1016/0550-3213(81)90361-8}
  {\bibfield  {journal} {\bibinfo  {journal} {Nuclear Physics B}\ }\textbf
  {\bibinfo {volume} {185}},\ \bibinfo {pages} {20 } (\bibinfo {year}
  {1981}{\natexlab{a}})}\BibitemShut {NoStop}%
\bibitem [{\citenamefont {Nielsen}\ and\ \citenamefont
  {Ninomiya}(1981{\natexlab{b}})}]{Nielsen81NPB2}%
  \BibitemOpen
  \bibfield  {author} {\bibinfo {author} {\bibfnamefont {H.}~\bibnamefont
  {Nielsen}}\ and\ \bibinfo {author} {\bibfnamefont {M.}~\bibnamefont
  {Ninomiya}},\ }\href {\doibase https://doi.org/10.1016/0550-3213(81)90524-1}
  {\bibfield  {journal} {\bibinfo  {journal} {Nuclear Physics B}\ }\textbf
  {\bibinfo {volume} {193}},\ \bibinfo {pages} {173 } (\bibinfo {year}
  {1981}{\natexlab{b}})}\BibitemShut {NoStop}%
\bibitem [{\citenamefont {Nielsen}\ and\ \citenamefont
  {Ninomiya}(1981{\natexlab{c}})}]{Nielsen81PLB}%
  \BibitemOpen
  \bibfield  {author} {\bibinfo {author} {\bibfnamefont {H.}~\bibnamefont
  {Nielsen}}\ and\ \bibinfo {author} {\bibfnamefont {M.}~\bibnamefont
  {Ninomiya}},\ }\href {\doibase https://doi.org/10.1016/0370-2693(81)91026-1}
  {\bibfield  {journal} {\bibinfo  {journal} {Physics Letters B}\ }\textbf
  {\bibinfo {volume} {105}},\ \bibinfo {pages} {219 } (\bibinfo {year}
  {1981}{\natexlab{c}})}\BibitemShut {NoStop}%
\bibitem [{\citenamefont {Alavirad}\ and\ \citenamefont
  {Sau}(2016)}]{Alavirad16}%
  \BibitemOpen
  \bibfield  {author} {\bibinfo {author} {\bibfnamefont {Y.}~\bibnamefont
  {Alavirad}}\ and\ \bibinfo {author} {\bibfnamefont {J.~D.}\ \bibnamefont
  {Sau}},\ }\href {\doibase 10.1103/PhysRevB.94.115160} {\bibfield  {journal}
  {\bibinfo  {journal} {Phys. Rev. B}\ }\textbf {\bibinfo {volume} {94}},\
  \bibinfo {pages} {115160} (\bibinfo {year} {2016})}\BibitemShut {NoStop}%
\bibitem [{\citenamefont {Dantas}\ \emph {et~al.}(2018)\citenamefont {Dantas},
  \citenamefont {Pe{\~n}a-Benitez}, \citenamefont {Roy},\ and\ \citenamefont
  {Sur{\'o}wka}}]{Dantas18}%
  \BibitemOpen
  \bibfield  {author} {\bibinfo {author} {\bibfnamefont {R.}~\bibnamefont
  {Dantas}}, \bibinfo {author} {\bibfnamefont {F.}~\bibnamefont
  {Pe{\~n}a-Benitez}}, \bibinfo {author} {\bibfnamefont {B.}~\bibnamefont
  {Roy}}, \ and\ \bibinfo {author} {\bibfnamefont {P.}~\bibnamefont
  {Sur{\'o}wka}},\ }\href@noop {} {\bibfield  {journal} {\bibinfo  {journal}
  {arXiv preprint arXiv:1802.07733}\ } (\bibinfo {year} {2018})}\BibitemShut
  {NoStop}%
\bibitem [{\citenamefont {Son}\ and\ \citenamefont {Spivak}(2013)}]{Son13}%
  \BibitemOpen
  \bibfield  {author} {\bibinfo {author} {\bibfnamefont {D.~T.}\ \bibnamefont
  {Son}}\ and\ \bibinfo {author} {\bibfnamefont {B.~Z.}\ \bibnamefont
  {Spivak}},\ }\href {\doibase 10.1103/PhysRevB.88.104412} {\bibfield
  {journal} {\bibinfo  {journal} {Phys. Rev. B}\ }\textbf {\bibinfo {volume}
  {88}},\ \bibinfo {pages} {104412} (\bibinfo {year} {2013})}\BibitemShut
  {NoStop}%
\bibitem [{\citenamefont {Xiong}\ \emph {et~al.}(2015)\citenamefont {Xiong},
  \citenamefont {Kushwaha}, \citenamefont {Liang}, \citenamefont {Krizan},
  \citenamefont {Hirschberger}, \citenamefont {Wang}, \citenamefont {Cava},\
  and\ \citenamefont {Ong}}]{xiong2015evidence}%
  \BibitemOpen
  \bibfield  {author} {\bibinfo {author} {\bibfnamefont {J.}~\bibnamefont
  {Xiong}}, \bibinfo {author} {\bibfnamefont {S.~K.}\ \bibnamefont {Kushwaha}},
  \bibinfo {author} {\bibfnamefont {T.}~\bibnamefont {Liang}}, \bibinfo
  {author} {\bibfnamefont {J.~W.}\ \bibnamefont {Krizan}}, \bibinfo {author}
  {\bibfnamefont {M.}~\bibnamefont {Hirschberger}}, \bibinfo {author}
  {\bibfnamefont {W.}~\bibnamefont {Wang}}, \bibinfo {author} {\bibfnamefont
  {R.}~\bibnamefont {Cava}}, \ and\ \bibinfo {author} {\bibfnamefont
  {N.}~\bibnamefont {Ong}},\ }\href@noop {} {\bibfield  {journal} {\bibinfo
  {journal} {Science}\ }\textbf {\bibinfo {volume} {350}},\ \bibinfo {pages}
  {413} (\bibinfo {year} {2015})}\BibitemShut {NoStop}%
\bibitem [{\citenamefont {Liang}\ \emph {et~al.}(2015)\citenamefont {Liang},
  \citenamefont {Gibson}, \citenamefont {Ali}, \citenamefont {Liu},
  \citenamefont {Cava},\ and\ \citenamefont {Ong}}]{liang2015ultrahigh}%
  \BibitemOpen
  \bibfield  {author} {\bibinfo {author} {\bibfnamefont {T.}~\bibnamefont
  {Liang}}, \bibinfo {author} {\bibfnamefont {Q.}~\bibnamefont {Gibson}},
  \bibinfo {author} {\bibfnamefont {M.~N.}\ \bibnamefont {Ali}}, \bibinfo
  {author} {\bibfnamefont {M.}~\bibnamefont {Liu}}, \bibinfo {author}
  {\bibfnamefont {R.}~\bibnamefont {Cava}}, \ and\ \bibinfo {author}
  {\bibfnamefont {N.}~\bibnamefont {Ong}},\ }\href@noop {} {\bibfield
  {journal} {\bibinfo  {journal} {Nature materials}\ }\textbf {\bibinfo
  {volume} {14}},\ \bibinfo {pages} {280} (\bibinfo {year} {2015})}\BibitemShut
  {NoStop}%
\bibitem [{\citenamefont {Huang}\ \emph {et~al.}(2015)\citenamefont {Huang},
  \citenamefont {Zhao}, \citenamefont {Long}, \citenamefont {Wang},
  \citenamefont {Chen}, \citenamefont {Yang}, \citenamefont {Liang},
  \citenamefont {Xue}, \citenamefont {Weng}, \citenamefont {Fang},
  \citenamefont {Dai},\ and\ \citenamefont {Chen}}]{Huang15}%
  \BibitemOpen
  \bibfield  {author} {\bibinfo {author} {\bibfnamefont {X.}~\bibnamefont
  {Huang}}, \bibinfo {author} {\bibfnamefont {L.}~\bibnamefont {Zhao}},
  \bibinfo {author} {\bibfnamefont {Y.}~\bibnamefont {Long}}, \bibinfo {author}
  {\bibfnamefont {P.}~\bibnamefont {Wang}}, \bibinfo {author} {\bibfnamefont
  {D.}~\bibnamefont {Chen}}, \bibinfo {author} {\bibfnamefont {Z.}~\bibnamefont
  {Yang}}, \bibinfo {author} {\bibfnamefont {H.}~\bibnamefont {Liang}},
  \bibinfo {author} {\bibfnamefont {M.}~\bibnamefont {Xue}}, \bibinfo {author}
  {\bibfnamefont {H.}~\bibnamefont {Weng}}, \bibinfo {author} {\bibfnamefont
  {Z.}~\bibnamefont {Fang}}, \bibinfo {author} {\bibfnamefont {X.}~\bibnamefont
  {Dai}}, \ and\ \bibinfo {author} {\bibfnamefont {G.}~\bibnamefont {Chen}},\
  }\href {\doibase 10.1103/PhysRevX.5.031023} {\bibfield  {journal} {\bibinfo
  {journal} {Phys. Rev. X}\ }\textbf {\bibinfo {volume} {5}},\ \bibinfo {pages}
  {031023} (\bibinfo {year} {2015})}\BibitemShut {NoStop}%
\bibitem [{\citenamefont {Zhang}\ \emph {et~al.}(2016)\citenamefont {Zhang},
  \citenamefont {Xu}, \citenamefont {Belopolski}, \citenamefont {Yuan},
  \citenamefont {Lin}, \citenamefont {Tong}, \citenamefont {Bian},
  \citenamefont {Alidoust}, \citenamefont {Lee}, \citenamefont {Huang} \emph
  {et~al.}}]{Zhang16}%
  \BibitemOpen
  \bibfield  {author} {\bibinfo {author} {\bibfnamefont {C.-L.}\ \bibnamefont
  {Zhang}}, \bibinfo {author} {\bibfnamefont {S.-Y.}\ \bibnamefont {Xu}},
  \bibinfo {author} {\bibfnamefont {I.}~\bibnamefont {Belopolski}}, \bibinfo
  {author} {\bibfnamefont {Z.}~\bibnamefont {Yuan}}, \bibinfo {author}
  {\bibfnamefont {Z.}~\bibnamefont {Lin}}, \bibinfo {author} {\bibfnamefont
  {B.}~\bibnamefont {Tong}}, \bibinfo {author} {\bibfnamefont {G.}~\bibnamefont
  {Bian}}, \bibinfo {author} {\bibfnamefont {N.}~\bibnamefont {Alidoust}},
  \bibinfo {author} {\bibfnamefont {C.-C.}\ \bibnamefont {Lee}}, \bibinfo
  {author} {\bibfnamefont {S.-M.}\ \bibnamefont {Huang}},  \emph {et~al.},\
  }\href@noop {} {\bibfield  {journal} {\bibinfo  {journal} {Nature
  communications}\ }\textbf {\bibinfo {volume} {7}},\ \bibinfo {pages} {10735}
  (\bibinfo {year} {2016})}\BibitemShut {NoStop}%
\bibitem [{\citenamefont {Yang}\ \emph
  {et~al.}(2015{\natexlab{b}})\citenamefont {Yang}, \citenamefont {Liu},
  \citenamefont {Wang}, \citenamefont {Zheng},\ and\ \citenamefont
  {Xu}}]{YangX15}%
  \BibitemOpen
  \bibfield  {author} {\bibinfo {author} {\bibfnamefont {X.}~\bibnamefont
  {Yang}}, \bibinfo {author} {\bibfnamefont {Y.}~\bibnamefont {Liu}}, \bibinfo
  {author} {\bibfnamefont {Z.}~\bibnamefont {Wang}}, \bibinfo {author}
  {\bibfnamefont {Y.}~\bibnamefont {Zheng}}, \ and\ \bibinfo {author}
  {\bibfnamefont {Z.-a.}\ \bibnamefont {Xu}},\ }\href@noop {} {\bibfield
  {journal} {\bibinfo  {journal} {arXiv preprint arXiv:1506.03190}\ } (\bibinfo
  {year} {2015}{\natexlab{b}})}\BibitemShut {NoStop}%
\bibitem [{\citenamefont {Du}\ \emph {et~al.}(2016)\citenamefont {Du},
  \citenamefont {Wang}, \citenamefont {Chen}, \citenamefont {Mao},
  \citenamefont {Khan}, \citenamefont {Xu}, \citenamefont {Zhou}, \citenamefont
  {Zhang}, \citenamefont {Yang}, \citenamefont {Chen}, \citenamefont {Feng},\
  and\ \citenamefont {Fang}}]{Du16}%
  \BibitemOpen
  \bibfield  {author} {\bibinfo {author} {\bibfnamefont {J.}~\bibnamefont
  {Du}}, \bibinfo {author} {\bibfnamefont {H.}~\bibnamefont {Wang}}, \bibinfo
  {author} {\bibfnamefont {Q.}~\bibnamefont {Chen}}, \bibinfo {author}
  {\bibfnamefont {Q.}~\bibnamefont {Mao}}, \bibinfo {author} {\bibfnamefont
  {R.}~\bibnamefont {Khan}}, \bibinfo {author} {\bibfnamefont {B.}~\bibnamefont
  {Xu}}, \bibinfo {author} {\bibfnamefont {Y.}~\bibnamefont {Zhou}}, \bibinfo
  {author} {\bibfnamefont {Y.}~\bibnamefont {Zhang}}, \bibinfo {author}
  {\bibfnamefont {J.}~\bibnamefont {Yang}}, \bibinfo {author} {\bibfnamefont
  {B.}~\bibnamefont {Chen}}, \bibinfo {author} {\bibfnamefont {C.}~\bibnamefont
  {Feng}}, \ and\ \bibinfo {author} {\bibfnamefont {M.}~\bibnamefont {Fang}},\
  }\href {\doibase 10.1007/s11433-016-5798-4} {\bibfield  {journal} {\bibinfo
  {journal} {Science China Physics, Mechanics {\&} Astronomy}\ }\textbf
  {\bibinfo {volume} {59}},\ \bibinfo {pages} {657406} (\bibinfo {year}
  {2016})}\BibitemShut {NoStop}%
\bibitem [{\citenamefont {Shekhar}\ \emph {et~al.}(2015)\citenamefont
  {Shekhar}, \citenamefont {Nayak}, \citenamefont {Sun}, \citenamefont
  {Schmidt}, \citenamefont {Nicklas}, \citenamefont {Leermakers}, \citenamefont
  {Zeitler}, \citenamefont {Skourski}, \citenamefont {Wosnitza}, \citenamefont
  {Liu} \emph {et~al.}}]{Shekhar15}%
  \BibitemOpen
  \bibfield  {author} {\bibinfo {author} {\bibfnamefont {C.}~\bibnamefont
  {Shekhar}}, \bibinfo {author} {\bibfnamefont {A.~K.}\ \bibnamefont {Nayak}},
  \bibinfo {author} {\bibfnamefont {Y.}~\bibnamefont {Sun}}, \bibinfo {author}
  {\bibfnamefont {M.}~\bibnamefont {Schmidt}}, \bibinfo {author} {\bibfnamefont
  {M.}~\bibnamefont {Nicklas}}, \bibinfo {author} {\bibfnamefont
  {I.}~\bibnamefont {Leermakers}}, \bibinfo {author} {\bibfnamefont
  {U.}~\bibnamefont {Zeitler}}, \bibinfo {author} {\bibfnamefont
  {Y.}~\bibnamefont {Skourski}}, \bibinfo {author} {\bibfnamefont
  {J.}~\bibnamefont {Wosnitza}}, \bibinfo {author} {\bibfnamefont
  {Z.}~\bibnamefont {Liu}},  \emph {et~al.},\ }\href@noop {} {\bibfield
  {journal} {\bibinfo  {journal} {Nature Physics}\ }\textbf {\bibinfo {volume}
  {11}},\ \bibinfo {pages} {645} (\bibinfo {year} {2015})}\BibitemShut
  {NoStop}%
\bibitem [{\citenamefont {Wang}\ \emph {et~al.}(2016)\citenamefont {Wang},
  \citenamefont {Zheng}, \citenamefont {Shen}, \citenamefont {Lu},
  \citenamefont {Fang}, \citenamefont {Sheng}, \citenamefont {Zhou},
  \citenamefont {Yang}, \citenamefont {Li}, \citenamefont {Feng},\ and\
  \citenamefont {Xu}}]{Wang16}%
  \BibitemOpen
  \bibfield  {author} {\bibinfo {author} {\bibfnamefont {Z.}~\bibnamefont
  {Wang}}, \bibinfo {author} {\bibfnamefont {Y.}~\bibnamefont {Zheng}},
  \bibinfo {author} {\bibfnamefont {Z.}~\bibnamefont {Shen}}, \bibinfo {author}
  {\bibfnamefont {Y.}~\bibnamefont {Lu}}, \bibinfo {author} {\bibfnamefont
  {H.}~\bibnamefont {Fang}}, \bibinfo {author} {\bibfnamefont {F.}~\bibnamefont
  {Sheng}}, \bibinfo {author} {\bibfnamefont {Y.}~\bibnamefont {Zhou}},
  \bibinfo {author} {\bibfnamefont {X.}~\bibnamefont {Yang}}, \bibinfo {author}
  {\bibfnamefont {Y.}~\bibnamefont {Li}}, \bibinfo {author} {\bibfnamefont
  {C.}~\bibnamefont {Feng}}, \ and\ \bibinfo {author} {\bibfnamefont {Z.-A.}\
  \bibnamefont {Xu}},\ }\href {\doibase 10.1103/PhysRevB.93.121112} {\bibfield
  {journal} {\bibinfo  {journal} {Phys. Rev. B}\ }\textbf {\bibinfo {volume}
  {93}},\ \bibinfo {pages} {121112} (\bibinfo {year} {2016})}\BibitemShut
  {NoStop}%
\bibitem [{\citenamefont {Nandkishore}\ \emph {et~al.}(2014)\citenamefont
  {Nandkishore}, \citenamefont {Huse},\ and\ \citenamefont
  {Sondhi}}]{Nandkishore14}%
  \BibitemOpen
  \bibfield  {author} {\bibinfo {author} {\bibfnamefont {R.}~\bibnamefont
  {Nandkishore}}, \bibinfo {author} {\bibfnamefont {D.~A.}\ \bibnamefont
  {Huse}}, \ and\ \bibinfo {author} {\bibfnamefont {S.~L.}\ \bibnamefont
  {Sondhi}},\ }\href {\doibase 10.1103/PhysRevB.89.245110} {\bibfield
  {journal} {\bibinfo  {journal} {Phys. Rev. B}\ }\textbf {\bibinfo {volume}
  {89}},\ \bibinfo {pages} {245110} (\bibinfo {year} {2014})}\BibitemShut
  {NoStop}%
\bibitem [{\citenamefont {Adam}\ \emph {et~al.}(2007)\citenamefont {Adam},
  \citenamefont {Hwang}, \citenamefont {Galitski},\ and\ \citenamefont
  {Sarma}}]{adam2007self}%
  \BibitemOpen
  \bibfield  {author} {\bibinfo {author} {\bibfnamefont {S.}~\bibnamefont
  {Adam}}, \bibinfo {author} {\bibfnamefont {E.}~\bibnamefont {Hwang}},
  \bibinfo {author} {\bibfnamefont {V.}~\bibnamefont {Galitski}}, \ and\
  \bibinfo {author} {\bibfnamefont {S.~D.}\ \bibnamefont {Sarma}},\ }\href@noop
  {} {\bibfield  {journal} {\bibinfo  {journal} {Proceedings of the National
  Academy of Sciences}\ }\textbf {\bibinfo {volume} {104}},\ \bibinfo {pages}
  {18392} (\bibinfo {year} {2007})}\BibitemShut {NoStop}%
\bibitem [{\citenamefont {Skinner}(2014)}]{Skinner-2014}%
  \BibitemOpen
  \bibfield  {author} {\bibinfo {author} {\bibfnamefont {B.}~\bibnamefont
  {Skinner}},\ }\href {\doibase 10.1103/PhysRevB.90.060202} {\bibfield
  {journal} {\bibinfo  {journal} {Phys. Rev. B}\ }\textbf {\bibinfo {volume}
  {90}},\ \bibinfo {pages} {060202} (\bibinfo {year} {2014})}\BibitemShut
  {NoStop}%
\bibitem [{\citenamefont {Nielsen}\ and\ \citenamefont
  {Ninomiya}(1983)}]{Nielsen83}%
  \BibitemOpen
  \bibfield  {author} {\bibinfo {author} {\bibfnamefont {H.}~\bibnamefont
  {Nielsen}}\ and\ \bibinfo {author} {\bibfnamefont {M.}~\bibnamefont
  {Ninomiya}},\ }\href {\doibase https://doi.org/10.1016/0370-2693(83)91529-0}
  {\bibfield  {journal} {\bibinfo  {journal} {Physics Letters B}\ }\textbf
  {\bibinfo {volume} {130}},\ \bibinfo {pages} {389 } (\bibinfo {year}
  {1983})}\BibitemShut {NoStop}%
\bibitem [{\citenamefont {Goswami}\ \emph {et~al.}(2015)\citenamefont
  {Goswami}, \citenamefont {Pixley},\ and\ \citenamefont
  {Das~Sarma}}]{Goswami15}%
  \BibitemOpen
  \bibfield  {author} {\bibinfo {author} {\bibfnamefont {P.}~\bibnamefont
  {Goswami}}, \bibinfo {author} {\bibfnamefont {J.~H.}\ \bibnamefont {Pixley}},
  \ and\ \bibinfo {author} {\bibfnamefont {S.}~\bibnamefont {Das~Sarma}},\
  }\href {\doibase 10.1103/PhysRevB.92.075205} {\bibfield  {journal} {\bibinfo
  {journal} {Phys. Rev. B}\ }\textbf {\bibinfo {volume} {92}},\ \bibinfo
  {pages} {075205} (\bibinfo {year} {2015})}\BibitemShut {NoStop}%
\bibitem [{\citenamefont {Biswas}\ and\ \citenamefont {Ryu}(2014)}]{Biswas14}%
  \BibitemOpen
  \bibfield  {author} {\bibinfo {author} {\bibfnamefont {R.~R.}\ \bibnamefont
  {Biswas}}\ and\ \bibinfo {author} {\bibfnamefont {S.}~\bibnamefont {Ryu}},\
  }\href {\doibase 10.1103/PhysRevB.89.014205} {\bibfield  {journal} {\bibinfo
  {journal} {Phys. Rev. B}\ }\textbf {\bibinfo {volume} {89}},\ \bibinfo
  {pages} {014205} (\bibinfo {year} {2014})}\BibitemShut {NoStop}%
\bibitem [{\citenamefont {Fradkin}(1986)}]{Fradkin-1986}%
  \BibitemOpen
  \bibfield  {author} {\bibinfo {author} {\bibfnamefont {E.}~\bibnamefont
  {Fradkin}},\ }\href@noop {} {\bibfield  {journal} {\bibinfo  {journal} {Phys.
  Rev. B}\ }\textbf {\bibinfo {volume} {33}},\ \bibinfo {pages} {3263}
  (\bibinfo {year} {1986})}\BibitemShut {NoStop}%
\bibitem [{\citenamefont {Goswami}\ and\ \citenamefont
  {Chakravarty}(2011)}]{Goswami-2011}%
  \BibitemOpen
  \bibfield  {author} {\bibinfo {author} {\bibfnamefont {P.}~\bibnamefont
  {Goswami}}\ and\ \bibinfo {author} {\bibfnamefont {S.}~\bibnamefont
  {Chakravarty}},\ }\href@noop {} {\bibfield  {journal} {\bibinfo  {journal}
  {Phys. Rev. Lett.}\ }\textbf {\bibinfo {volume} {107}},\ \bibinfo {pages}
  {196803} (\bibinfo {year} {2011})}\BibitemShut {NoStop}%
\bibitem [{\citenamefont {Kobayashi}\ \emph {et~al.}(2014)\citenamefont
  {Kobayashi}, \citenamefont {Ohtsuki}, \citenamefont {Imura},\ and\
  \citenamefont {Herbut}}]{Kobayashi-2014}%
  \BibitemOpen
  \bibfield  {author} {\bibinfo {author} {\bibfnamefont {K.}~\bibnamefont
  {Kobayashi}}, \bibinfo {author} {\bibfnamefont {T.}~\bibnamefont {Ohtsuki}},
  \bibinfo {author} {\bibfnamefont {K.-I.}\ \bibnamefont {Imura}}, \ and\
  \bibinfo {author} {\bibfnamefont {I.~F.}\ \bibnamefont {Herbut}},\
  }\href@noop {} {\bibfield  {journal} {\bibinfo  {journal} {Phys. Rev. Lett.}\
  }\textbf {\bibinfo {volume} {112}},\ \bibinfo {pages} {016402} (\bibinfo
  {year} {2014})}\BibitemShut {NoStop}%
\bibitem [{\citenamefont {{Moon}}\ and\ \citenamefont {{Kim}}(2014)}]{Moon14}%
  \BibitemOpen
  \bibfield  {author} {\bibinfo {author} {\bibfnamefont {E.-G.}\ \bibnamefont
  {{Moon}}}\ and\ \bibinfo {author} {\bibfnamefont {Y.~B.}\ \bibnamefont
  {{Kim}}},\ }\href@noop {} {\bibfield  {journal} {\bibinfo  {journal} {ArXiv
  e-prints}\ } (\bibinfo {year} {2014})},\ \Eprint
  {http://arxiv.org/abs/1409.0573} {arXiv:1409.0573 [cond-mat.str-el]}
  \BibitemShut {NoStop}%
\bibitem [{\citenamefont {Roy}\ and\ \citenamefont
  {Das~Sarma}(2014)}]{Bitan-2014}%
  \BibitemOpen
  \bibfield  {author} {\bibinfo {author} {\bibfnamefont {B.}~\bibnamefont
  {Roy}}\ and\ \bibinfo {author} {\bibfnamefont {S.}~\bibnamefont
  {Das~Sarma}},\ }\href@noop {} {\bibfield  {journal} {\bibinfo  {journal}
  {Phys. Rev. B}\ }\textbf {\bibinfo {volume} {90}},\ \bibinfo {pages} {241112}
  (\bibinfo {year} {2014})}\BibitemShut {NoStop}%
\bibitem [{\citenamefont {Roy}\ and\ \citenamefont
  {Das~Sarma}(2016)}]{Bitan-2016}%
  \BibitemOpen
  \bibfield  {author} {\bibinfo {author} {\bibfnamefont {B.}~\bibnamefont
  {Roy}}\ and\ \bibinfo {author} {\bibfnamefont {S.}~\bibnamefont
  {Das~Sarma}},\ }\href@noop {} {\bibfield  {journal} {\bibinfo  {journal}
  {Phys. Rev. B}\ }\textbf {\bibinfo {volume} {93}},\ \bibinfo {pages} {119911}
  (\bibinfo {year} {2016})}\BibitemShut {NoStop}%
\bibitem [{\citenamefont {Sbierski}\ \emph {et~al.}(2014)\citenamefont
  {Sbierski}, \citenamefont {Pohl}, \citenamefont {Bergholtz},\ and\
  \citenamefont {Brouwer}}]{Brouwer-2014}%
  \BibitemOpen
  \bibfield  {author} {\bibinfo {author} {\bibfnamefont {B.}~\bibnamefont
  {Sbierski}}, \bibinfo {author} {\bibfnamefont {G.}~\bibnamefont {Pohl}},
  \bibinfo {author} {\bibfnamefont {E.~J.}\ \bibnamefont {Bergholtz}}, \ and\
  \bibinfo {author} {\bibfnamefont {P.~W.}\ \bibnamefont {Brouwer}},\
  }\href@noop {} {\bibfield  {journal} {\bibinfo  {journal} {Phys. Rev. Lett.}\
  }\textbf {\bibinfo {volume} {113}},\ \bibinfo {pages} {026602} (\bibinfo
  {year} {2014})}\BibitemShut {NoStop}%
\bibitem [{\citenamefont {Altland}\ and\ \citenamefont
  {Bagrets}(2015)}]{Altland-2015}%
  \BibitemOpen
  \bibfield  {author} {\bibinfo {author} {\bibfnamefont {A.}~\bibnamefont
  {Altland}}\ and\ \bibinfo {author} {\bibfnamefont {D.}~\bibnamefont
  {Bagrets}},\ }\href@noop {} {\bibfield  {journal} {\bibinfo  {journal} {Phys.
  Rev. Lett.}\ }\textbf {\bibinfo {volume} {114}},\ \bibinfo {pages} {257201}
  (\bibinfo {year} {2015})}\BibitemShut {NoStop}%
\bibitem [{\citenamefont {G\"arttner}\ \emph {et~al.}(2015)\citenamefont
  {G\"arttner}, \citenamefont {Syzranov}, \citenamefont {Rey}, \citenamefont
  {Gurarie},\ and\ \citenamefont {Radzihovsky}}]{Garttner-2015}%
  \BibitemOpen
  \bibfield  {author} {\bibinfo {author} {\bibfnamefont {M.}~\bibnamefont
  {G\"arttner}}, \bibinfo {author} {\bibfnamefont {S.~V.}\ \bibnamefont
  {Syzranov}}, \bibinfo {author} {\bibfnamefont {A.~M.}\ \bibnamefont {Rey}},
  \bibinfo {author} {\bibfnamefont {V.}~\bibnamefont {Gurarie}}, \ and\
  \bibinfo {author} {\bibfnamefont {L.}~\bibnamefont {Radzihovsky}},\
  }\href@noop {} {\bibfield  {journal} {\bibinfo  {journal} {Phys. Rev. B}\
  }\textbf {\bibinfo {volume} {92}},\ \bibinfo {pages} {041406} (\bibinfo
  {year} {2015})}\BibitemShut {NoStop}%
\bibitem [{\citenamefont {Pixley}\ \emph {et~al.}(2015)\citenamefont {Pixley},
  \citenamefont {Goswami},\ and\ \citenamefont {Das~Sarma}}]{Pixley15}%
  \BibitemOpen
  \bibfield  {author} {\bibinfo {author} {\bibfnamefont {J.~H.}\ \bibnamefont
  {Pixley}}, \bibinfo {author} {\bibfnamefont {P.}~\bibnamefont {Goswami}}, \
  and\ \bibinfo {author} {\bibfnamefont {S.}~\bibnamefont {Das~Sarma}},\ }\href
  {\doibase 10.1103/PhysRevLett.115.076601} {\bibfield  {journal} {\bibinfo
  {journal} {Phys. Rev. Lett.}\ }\textbf {\bibinfo {volume} {115}},\ \bibinfo
  {pages} {076601} (\bibinfo {year} {2015})}\BibitemShut {NoStop}%
\bibitem [{\citenamefont {Sbierski}\ \emph {et~al.}(2015)\citenamefont
  {Sbierski}, \citenamefont {Bergholtz},\ and\ \citenamefont
  {Brouwer}}]{Sbierski-2015}%
  \BibitemOpen
  \bibfield  {author} {\bibinfo {author} {\bibfnamefont {B.}~\bibnamefont
  {Sbierski}}, \bibinfo {author} {\bibfnamefont {E.~J.}\ \bibnamefont
  {Bergholtz}}, \ and\ \bibinfo {author} {\bibfnamefont {P.~W.}\ \bibnamefont
  {Brouwer}},\ }\href@noop {} {\bibfield  {journal} {\bibinfo  {journal} {Phys.
  Rev. B}\ }\textbf {\bibinfo {volume} {92}},\ \bibinfo {pages} {115145}
  (\bibinfo {year} {2015})}\BibitemShut {NoStop}%
\bibitem [{\citenamefont {Syzranov}\ \emph
  {et~al.}(2015{\natexlab{a}})\citenamefont {Syzranov}, \citenamefont
  {Gurarie},\ and\ \citenamefont {Radzihovsky}}]{Leo-2015}%
  \BibitemOpen
  \bibfield  {author} {\bibinfo {author} {\bibfnamefont {S.~V.}\ \bibnamefont
  {Syzranov}}, \bibinfo {author} {\bibfnamefont {V.}~\bibnamefont {Gurarie}}, \
  and\ \bibinfo {author} {\bibfnamefont {L.}~\bibnamefont {Radzihovsky}},\
  }\href@noop {} {\bibfield  {journal} {\bibinfo  {journal} {Phys. Rev. B}\
  }\textbf {\bibinfo {volume} {91}},\ \bibinfo {pages} {035133} (\bibinfo
  {year} {2015}{\natexlab{a}})}\BibitemShut {NoStop}%
\bibitem [{\citenamefont {Syzranov}\ \emph
  {et~al.}(2015{\natexlab{b}})\citenamefont {Syzranov}, \citenamefont
  {Radzihovsky},\ and\ \citenamefont {Gurarie}}]{Sergey-2015}%
  \BibitemOpen
  \bibfield  {author} {\bibinfo {author} {\bibfnamefont {S.~V.}\ \bibnamefont
  {Syzranov}}, \bibinfo {author} {\bibfnamefont {L.}~\bibnamefont
  {Radzihovsky}}, \ and\ \bibinfo {author} {\bibfnamefont {V.}~\bibnamefont
  {Gurarie}},\ }\href@noop {} {\bibfield  {journal} {\bibinfo  {journal} {Phys.
  Rev. Lett.}\ }\textbf {\bibinfo {volume} {114}},\ \bibinfo {pages} {166601}
  (\bibinfo {year} {2015}{\natexlab{b}})}\BibitemShut {NoStop}%
\bibitem [{\citenamefont {Altland}\ and\ \citenamefont
  {Bagrets}(2016)}]{Altland16}%
  \BibitemOpen
  \bibfield  {author} {\bibinfo {author} {\bibfnamefont {A.}~\bibnamefont
  {Altland}}\ and\ \bibinfo {author} {\bibfnamefont {D.}~\bibnamefont
  {Bagrets}},\ }\href {\doibase 10.1103/PhysRevB.93.075113} {\bibfield
  {journal} {\bibinfo  {journal} {Phys. Rev. B}\ }\textbf {\bibinfo {volume}
  {93}},\ \bibinfo {pages} {075113} (\bibinfo {year} {2016})}\BibitemShut
  {NoStop}%
\bibitem [{\citenamefont {Bera}\ \emph {et~al.}(2016)\citenamefont {Bera},
  \citenamefont {Sau},\ and\ \citenamefont {Roy}}]{Bera-2015}%
  \BibitemOpen
  \bibfield  {author} {\bibinfo {author} {\bibfnamefont {S.}~\bibnamefont
  {Bera}}, \bibinfo {author} {\bibfnamefont {J.~D.}\ \bibnamefont {Sau}}, \
  and\ \bibinfo {author} {\bibfnamefont {B.}~\bibnamefont {Roy}},\ }\href@noop
  {} {\bibfield  {journal} {\bibinfo  {journal} {Phys. Rev. B}\ }\textbf
  {\bibinfo {volume} {93}},\ \bibinfo {pages} {201302} (\bibinfo {year}
  {2016})}\BibitemShut {NoStop}%
\bibitem [{\citenamefont {Liu}\ \emph {et~al.}(2016)\citenamefont {Liu},
  \citenamefont {Ohtsuki},\ and\ \citenamefont {Shindou}}]{Liu-2015}%
  \BibitemOpen
  \bibfield  {author} {\bibinfo {author} {\bibfnamefont {S.}~\bibnamefont
  {Liu}}, \bibinfo {author} {\bibfnamefont {T.}~\bibnamefont {Ohtsuki}}, \ and\
  \bibinfo {author} {\bibfnamefont {R.}~\bibnamefont {Shindou}},\ }\href@noop
  {} {\bibfield  {journal} {\bibinfo  {journal} {Phys. Rev. Lett.}\ }\textbf
  {\bibinfo {volume} {116}},\ \bibinfo {pages} {066401} (\bibinfo {year}
  {2016})}\BibitemShut {NoStop}%
\bibitem [{\citenamefont {Louvet}\ \emph {et~al.}(2016)\citenamefont {Louvet},
  \citenamefont {Carpentier},\ and\ \citenamefont {Fedorenko}}]{Louvet-2016}%
  \BibitemOpen
  \bibfield  {author} {\bibinfo {author} {\bibfnamefont {T.}~\bibnamefont
  {Louvet}}, \bibinfo {author} {\bibfnamefont {D.}~\bibnamefont {Carpentier}},
  \ and\ \bibinfo {author} {\bibfnamefont {A.~A.}\ \bibnamefont {Fedorenko}},\
  }\href {\doibase 10.1103/PhysRevB.94.220201} {\bibfield  {journal} {\bibinfo
  {journal} {Phys. Rev. B}\ }\textbf {\bibinfo {volume} {94}},\ \bibinfo
  {pages} {220201} (\bibinfo {year} {2016})}\BibitemShut {NoStop}%
\bibitem [{\citenamefont {Pixley}\ \emph
  {et~al.}(2016{\natexlab{a}})\citenamefont {Pixley}, \citenamefont {Huse},\
  and\ \citenamefont {Das~Sarma}}]{PixleyPRX16}%
  \BibitemOpen
  \bibfield  {author} {\bibinfo {author} {\bibfnamefont {J.~H.}\ \bibnamefont
  {Pixley}}, \bibinfo {author} {\bibfnamefont {D.~A.}\ \bibnamefont {Huse}}, \
  and\ \bibinfo {author} {\bibfnamefont {S.}~\bibnamefont {Das~Sarma}},\ }\href
  {\doibase 10.1103/PhysRevX.6.021042} {\bibfield  {journal} {\bibinfo
  {journal} {Phys. Rev. X}\ }\textbf {\bibinfo {volume} {6}},\ \bibinfo {pages}
  {021042} (\bibinfo {year} {2016}{\natexlab{a}})}\BibitemShut {NoStop}%
\bibitem [{\citenamefont {Pixley}\ \emph
  {et~al.}(2016{\natexlab{b}})\citenamefont {Pixley}, \citenamefont {Goswami},\
  and\ \citenamefont {Das~Sarma}}]{PixleyPRB16}%
  \BibitemOpen
  \bibfield  {author} {\bibinfo {author} {\bibfnamefont {J.~H.}\ \bibnamefont
  {Pixley}}, \bibinfo {author} {\bibfnamefont {P.}~\bibnamefont {Goswami}}, \
  and\ \bibinfo {author} {\bibfnamefont {S.}~\bibnamefont {Das~Sarma}},\ }\href
  {\doibase 10.1103/PhysRevB.93.085103} {\bibfield  {journal} {\bibinfo
  {journal} {Phys. Rev. B}\ }\textbf {\bibinfo {volume} {93}},\ \bibinfo
  {pages} {085103} (\bibinfo {year} {2016}{\natexlab{b}})}\BibitemShut
  {NoStop}%
\bibitem [{\citenamefont {Pixley}\ \emph
  {et~al.}(2016{\natexlab{c}})\citenamefont {Pixley}, \citenamefont {Huse},\
  and\ \citenamefont {Das~Sarma}}]{Pixley2}%
  \BibitemOpen
  \bibfield  {author} {\bibinfo {author} {\bibfnamefont {J.~H.}\ \bibnamefont
  {Pixley}}, \bibinfo {author} {\bibfnamefont {D.~A.}\ \bibnamefont {Huse}}, \
  and\ \bibinfo {author} {\bibfnamefont {S.}~\bibnamefont {Das~Sarma}},\ }\href
  {\doibase 10.1103/PhysRevB.94.121107} {\bibfield  {journal} {\bibinfo
  {journal} {Phys. Rev. B}\ }\textbf {\bibinfo {volume} {94}},\ \bibinfo
  {pages} {121107} (\bibinfo {year} {2016}{\natexlab{c}})}\BibitemShut
  {NoStop}%
\bibitem [{\citenamefont {Shapourian}\ and\ \citenamefont
  {Hughes}(2016)}]{Shapourian-2015}%
  \BibitemOpen
  \bibfield  {author} {\bibinfo {author} {\bibfnamefont {H.}~\bibnamefont
  {Shapourian}}\ and\ \bibinfo {author} {\bibfnamefont {T.~L.}\ \bibnamefont
  {Hughes}},\ }\href@noop {} {\bibfield  {journal} {\bibinfo  {journal} {Phys.
  Rev. B}\ }\textbf {\bibinfo {volume} {93}},\ \bibinfo {pages} {075108}
  (\bibinfo {year} {2016})}\BibitemShut {NoStop}%
\bibitem [{\citenamefont {Syzranov}\ \emph {et~al.}(2016)\citenamefont
  {Syzranov}, \citenamefont {Ostrovsky}, \citenamefont {Gurarie},\ and\
  \citenamefont {Radzihovsky}}]{Sergey2-2015}%
  \BibitemOpen
  \bibfield  {author} {\bibinfo {author} {\bibfnamefont {S.~V.}\ \bibnamefont
  {Syzranov}}, \bibinfo {author} {\bibfnamefont {P.~M.}\ \bibnamefont
  {Ostrovsky}}, \bibinfo {author} {\bibfnamefont {V.}~\bibnamefont {Gurarie}},
  \ and\ \bibinfo {author} {\bibfnamefont {L.}~\bibnamefont {Radzihovsky}},\
  }\href@noop {} {\bibfield  {journal} {\bibinfo  {journal} {Phys. Rev. B}\
  }\textbf {\bibinfo {volume} {93}},\ \bibinfo {pages} {155113} (\bibinfo
  {year} {2016})}\BibitemShut {NoStop}%
\bibitem [{\citenamefont {Gurarie}(2017)}]{Guararie-2017}%
  \BibitemOpen
  \bibfield  {author} {\bibinfo {author} {\bibfnamefont {V.}~\bibnamefont
  {Gurarie}},\ }\href {\doibase 10.1103/PhysRevB.96.014205} {\bibfield
  {journal} {\bibinfo  {journal} {Phys. Rev. B}\ }\textbf {\bibinfo {volume}
  {96}},\ \bibinfo {pages} {014205} (\bibinfo {year} {2017})}\BibitemShut
  {NoStop}%
\bibitem [{\citenamefont {Pixley}\ \emph {et~al.}(2017)\citenamefont {Pixley},
  \citenamefont {Chou}, \citenamefont {Goswami}, \citenamefont {Huse},
  \citenamefont {Nandkishore}, \citenamefont {Radzihovsky},\ and\ \citenamefont
  {Das~Sarma}}]{Pixley17}%
  \BibitemOpen
  \bibfield  {author} {\bibinfo {author} {\bibfnamefont {J.~H.}\ \bibnamefont
  {Pixley}}, \bibinfo {author} {\bibfnamefont {Y.-Z.}\ \bibnamefont {Chou}},
  \bibinfo {author} {\bibfnamefont {P.}~\bibnamefont {Goswami}}, \bibinfo
  {author} {\bibfnamefont {D.~A.}\ \bibnamefont {Huse}}, \bibinfo {author}
  {\bibfnamefont {R.}~\bibnamefont {Nandkishore}}, \bibinfo {author}
  {\bibfnamefont {L.}~\bibnamefont {Radzihovsky}}, \ and\ \bibinfo {author}
  {\bibfnamefont {S.}~\bibnamefont {Das~Sarma}},\ }\href {\doibase
  10.1103/PhysRevB.95.235101} {\bibfield  {journal} {\bibinfo  {journal} {Phys.
  Rev. B}\ }\textbf {\bibinfo {volume} {95}},\ \bibinfo {pages} {235101}
  (\bibinfo {year} {2017})}\BibitemShut {NoStop}%
\bibitem [{\citenamefont {Sbierski}\ \emph {et~al.}(2017)\citenamefont
  {Sbierski}, \citenamefont {Madsen}, \citenamefont {Brouwer},\ and\
  \citenamefont {Karrasch}}]{Sbierski-2017}%
  \BibitemOpen
  \bibfield  {author} {\bibinfo {author} {\bibfnamefont {B.}~\bibnamefont
  {Sbierski}}, \bibinfo {author} {\bibfnamefont {K.~A.}\ \bibnamefont
  {Madsen}}, \bibinfo {author} {\bibfnamefont {P.~W.}\ \bibnamefont {Brouwer}},
  \ and\ \bibinfo {author} {\bibfnamefont {C.}~\bibnamefont {Karrasch}},\
  }\href {\doibase 10.1103/PhysRevB.96.064203} {\bibfield  {journal} {\bibinfo
  {journal} {Phys. Rev. B}\ }\textbf {\bibinfo {volume} {96}},\ \bibinfo
  {pages} {064203} (\bibinfo {year} {2017})}\BibitemShut {NoStop}%
\bibitem [{\citenamefont {Wilson}\ \emph {et~al.}(2017)\citenamefont {Wilson},
  \citenamefont {Pixley}, \citenamefont {Goswami},\ and\ \citenamefont
  {Das~Sarma}}]{Wilson17}%
  \BibitemOpen
  \bibfield  {author} {\bibinfo {author} {\bibfnamefont {J.~H.}\ \bibnamefont
  {Wilson}}, \bibinfo {author} {\bibfnamefont {J.~H.}\ \bibnamefont {Pixley}},
  \bibinfo {author} {\bibfnamefont {P.}~\bibnamefont {Goswami}}, \ and\
  \bibinfo {author} {\bibfnamefont {S.}~\bibnamefont {Das~Sarma}},\ }\href
  {\doibase 10.1103/PhysRevB.95.155122} {\bibfield  {journal} {\bibinfo
  {journal} {Phys. Rev. B}\ }\textbf {\bibinfo {volume} {95}},\ \bibinfo
  {pages} {155122} (\bibinfo {year} {2017})}\BibitemShut {NoStop}%
\bibitem [{\citenamefont {Wilson}\ \emph {et~al.}(2018)\citenamefont {Wilson},
  \citenamefont {Pixley}, \citenamefont {Huse}, \citenamefont {Refael},\ and\
  \citenamefont {Das~Sarma}}]{Wilson18}%
  \BibitemOpen
  \bibfield  {author} {\bibinfo {author} {\bibfnamefont {J.~H.}\ \bibnamefont
  {Wilson}}, \bibinfo {author} {\bibfnamefont {J.~H.}\ \bibnamefont {Pixley}},
  \bibinfo {author} {\bibfnamefont {D.~A.}\ \bibnamefont {Huse}}, \bibinfo
  {author} {\bibfnamefont {G.}~\bibnamefont {Refael}}, \ and\ \bibinfo {author}
  {\bibfnamefont {S.}~\bibnamefont {Das~Sarma}},\ }\href {\doibase
  10.1103/PhysRevB.97.235108} {\bibfield  {journal} {\bibinfo  {journal} {Phys.
  Rev. B}\ }\textbf {\bibinfo {volume} {97}},\ \bibinfo {pages} {235108}
  (\bibinfo {year} {2018})}\BibitemShut {NoStop}%
\bibitem [{\citenamefont {Syzranov}\ and\ \citenamefont
  {Radzihovsky}(2018)}]{Syzranov-2016}%
  \BibitemOpen
  \bibfield  {author} {\bibinfo {author} {\bibfnamefont {S.~V.}\ \bibnamefont
  {Syzranov}}\ and\ \bibinfo {author} {\bibfnamefont {L.}~\bibnamefont
  {Radzihovsky}},\ }\href {\doibase 10.1146/annurev-conmatphys-033117-054037}
  {\bibfield  {journal} {\bibinfo  {journal} {Annual Review of Condensed Matter
  Physics}\ }\textbf {\bibinfo {volume} {9}},\ \bibinfo {pages} {35} (\bibinfo
  {year} {2018})}\BibitemShut {NoStop}%
\bibitem [{\citenamefont {Haldane}(2014)}]{Haldane14}%
  \BibitemOpen
  \bibfield  {author} {\bibinfo {author} {\bibfnamefont {F.}~\bibnamefont
  {Haldane}},\ }\href@noop {} {\bibfield  {journal} {\bibinfo  {journal} {arXiv
  preprint arXiv:1401.0529}\ } (\bibinfo {year} {2014})}\BibitemShut {NoStop}%
\bibitem [{\citenamefont {Efetov}(1996)}]{Efetov96}%
  \BibitemOpen
  \bibfield  {author} {\bibinfo {author} {\bibfnamefont {K.}~\bibnamefont
  {Efetov}},\ }\href {\doibase 10.1017/CBO9780511573057} {\emph {\bibinfo
  {title} {Supersymmetry in Disorder and Chaos}}}\ (\bibinfo  {publisher}
  {Cambridge University Press},\ \bibinfo {year} {1996})\BibitemShut {NoStop}%
\bibitem [{\citenamefont {Simons}\ and\ \citenamefont
  {Altshuler}(1993{\natexlab{a}})}]{AltshulerPRL93}%
  \BibitemOpen
  \bibfield  {author} {\bibinfo {author} {\bibfnamefont {B.~D.}\ \bibnamefont
  {Simons}}\ and\ \bibinfo {author} {\bibfnamefont {B.~L.}\ \bibnamefont
  {Altshuler}},\ }\href {\doibase 10.1103/PhysRevLett.70.4063} {\bibfield
  {journal} {\bibinfo  {journal} {Phys. Rev. Lett.}\ }\textbf {\bibinfo
  {volume} {70}},\ \bibinfo {pages} {4063} (\bibinfo {year}
  {1993}{\natexlab{a}})}\BibitemShut {NoStop}%
\bibitem [{\citenamefont {Simons}\ and\ \citenamefont
  {Altshuler}(1993{\natexlab{b}})}]{AltshulerPRB93}%
  \BibitemOpen
  \bibfield  {author} {\bibinfo {author} {\bibfnamefont {B.~D.}\ \bibnamefont
  {Simons}}\ and\ \bibinfo {author} {\bibfnamefont {B.~L.}\ \bibnamefont
  {Altshuler}},\ }\href {\doibase 10.1103/PhysRevB.48.5422} {\bibfield
  {journal} {\bibinfo  {journal} {Phys. Rev. B}\ }\textbf {\bibinfo {volume}
  {48}},\ \bibinfo {pages} {5422} (\bibinfo {year}
  {1993}{\natexlab{b}})}\BibitemShut {NoStop}%
\bibitem [{\citenamefont {Ryu}\ \emph {et~al.}(2007)\citenamefont {Ryu},
  \citenamefont {Mudry}, \citenamefont {Obuse},\ and\ \citenamefont
  {Furusaki}}]{Ryu07}%
  \BibitemOpen
  \bibfield  {author} {\bibinfo {author} {\bibfnamefont {S.}~\bibnamefont
  {Ryu}}, \bibinfo {author} {\bibfnamefont {C.}~\bibnamefont {Mudry}}, \bibinfo
  {author} {\bibfnamefont {H.}~\bibnamefont {Obuse}}, \ and\ \bibinfo {author}
  {\bibfnamefont {A.}~\bibnamefont {Furusaki}},\ }\href {\doibase
  10.1103/PhysRevLett.99.116601} {\bibfield  {journal} {\bibinfo  {journal}
  {Phys. Rev. Lett.}\ }\textbf {\bibinfo {volume} {99}},\ \bibinfo {pages}
  {116601} (\bibinfo {year} {2007})}\BibitemShut {NoStop}%
\bibitem [{\citenamefont {Fu}\ and\ \citenamefont {Kane}(2012)}]{Fu12}%
  \BibitemOpen
  \bibfield  {author} {\bibinfo {author} {\bibfnamefont {L.}~\bibnamefont
  {Fu}}\ and\ \bibinfo {author} {\bibfnamefont {C.~L.}\ \bibnamefont {Kane}},\
  }\href {\doibase 10.1103/PhysRevLett.109.246605} {\bibfield  {journal}
  {\bibinfo  {journal} {Phys. Rev. Lett.}\ }\textbf {\bibinfo {volume} {109}},\
  \bibinfo {pages} {246605} (\bibinfo {year} {2012})}\BibitemShut {NoStop}%
\bibitem [{\citenamefont {Foster}\ and\ \citenamefont
  {Yuzbashyan}(2012)}]{Foster12}%
  \BibitemOpen
  \bibfield  {author} {\bibinfo {author} {\bibfnamefont {M.~S.}\ \bibnamefont
  {Foster}}\ and\ \bibinfo {author} {\bibfnamefont {E.~A.}\ \bibnamefont
  {Yuzbashyan}},\ }\href {\doibase 10.1103/PhysRevLett.109.246801} {\bibfield
  {journal} {\bibinfo  {journal} {Phys. Rev. Lett.}\ }\textbf {\bibinfo
  {volume} {109}},\ \bibinfo {pages} {246801} (\bibinfo {year}
  {2012})}\BibitemShut {NoStop}%
\bibitem [{\citenamefont {Fulga}\ \emph {et~al.}(2014)\citenamefont {Fulga},
  \citenamefont {van Heck}, \citenamefont {Edge},\ and\ \citenamefont
  {Akhmerov}}]{Fulga14}%
  \BibitemOpen
  \bibfield  {author} {\bibinfo {author} {\bibfnamefont {I.~C.}\ \bibnamefont
  {Fulga}}, \bibinfo {author} {\bibfnamefont {B.}~\bibnamefont {van Heck}},
  \bibinfo {author} {\bibfnamefont {J.~M.}\ \bibnamefont {Edge}}, \ and\
  \bibinfo {author} {\bibfnamefont {A.~R.}\ \bibnamefont {Akhmerov}},\ }\href
  {\doibase 10.1103/PhysRevB.89.155424} {\bibfield  {journal} {\bibinfo
  {journal} {Phys. Rev. B}\ }\textbf {\bibinfo {volume} {89}},\ \bibinfo
  {pages} {155424} (\bibinfo {year} {2014})}\BibitemShut {NoStop}%
\bibitem [{\citenamefont {Goswami}\ and\ \citenamefont
  {Tewari}(2013)}]{Goswami-2013}%
  \BibitemOpen
  \bibfield  {author} {\bibinfo {author} {\bibfnamefont {P.}~\bibnamefont
  {Goswami}}\ and\ \bibinfo {author} {\bibfnamefont {S.}~\bibnamefont
  {Tewari}},\ }\href@noop {} {\bibfield  {journal} {\bibinfo  {journal}
  {arXiv:1311.1506}\ } (\bibinfo {year} {2013})}\BibitemShut {NoStop}%
\bibitem [{\citenamefont {Vazifeh}\ and\ \citenamefont
  {Franz}(2013)}]{Vazifeh13}%
  \BibitemOpen
  \bibfield  {author} {\bibinfo {author} {\bibfnamefont {M.~M.}\ \bibnamefont
  {Vazifeh}}\ and\ \bibinfo {author} {\bibfnamefont {M.}~\bibnamefont
  {Franz}},\ }\href {\doibase 10.1103/PhysRevLett.111.027201} {\bibfield
  {journal} {\bibinfo  {journal} {Phys. Rev. Lett.}\ }\textbf {\bibinfo
  {volume} {111}},\ \bibinfo {pages} {027201} (\bibinfo {year}
  {2013})}\BibitemShut {NoStop}%
\bibitem [{\citenamefont {Ma}\ and\ \citenamefont {Pesin}(2015)}]{Ma15}%
  \BibitemOpen
  \bibfield  {author} {\bibinfo {author} {\bibfnamefont {J.}~\bibnamefont
  {Ma}}\ and\ \bibinfo {author} {\bibfnamefont {D.~A.}\ \bibnamefont {Pesin}},\
  }\href {\doibase 10.1103/PhysRevB.92.235205} {\bibfield  {journal} {\bibinfo
  {journal} {Phys. Rev. B}\ }\textbf {\bibinfo {volume} {92}},\ \bibinfo
  {pages} {235205} (\bibinfo {year} {2015})}\BibitemShut {NoStop}%
\bibitem [{\citenamefont {Zhong}\ \emph {et~al.}(2016)\citenamefont {Zhong},
  \citenamefont {Moore},\ and\ \citenamefont {Souza}}]{Zhong16}%
  \BibitemOpen
  \bibfield  {author} {\bibinfo {author} {\bibfnamefont {S.}~\bibnamefont
  {Zhong}}, \bibinfo {author} {\bibfnamefont {J.~E.}\ \bibnamefont {Moore}}, \
  and\ \bibinfo {author} {\bibfnamefont {I.}~\bibnamefont {Souza}},\ }\href
  {\doibase 10.1103/PhysRevLett.116.077201} {\bibfield  {journal} {\bibinfo
  {journal} {Phys. Rev. Lett.}\ }\textbf {\bibinfo {volume} {116}},\ \bibinfo
  {pages} {077201} (\bibinfo {year} {2016})}\BibitemShut {NoStop}%
\bibitem [{\citenamefont {Jian-Hui}\ \emph {et~al.}(2013)\citenamefont
  {Jian-Hui}, \citenamefont {Hua}, \citenamefont {Qian},\ and\ \citenamefont
  {Jun-Ren}}]{Zhou13}%
  \BibitemOpen
  \bibfield  {author} {\bibinfo {author} {\bibfnamefont {Z.}~\bibnamefont
  {Jian-Hui}}, \bibinfo {author} {\bibfnamefont {J.}~\bibnamefont {Hua}},
  \bibinfo {author} {\bibfnamefont {N.}~\bibnamefont {Qian}}, \ and\ \bibinfo
  {author} {\bibfnamefont {S.}~\bibnamefont {Jun-Ren}},\ }\href
  {http://stacks.iop.org/0256-307X/30/i=2/a=027101} {\bibfield  {journal}
  {\bibinfo  {journal} {Chinese Physics Letters}\ }\textbf {\bibinfo {volume}
  {30}},\ \bibinfo {pages} {027101} (\bibinfo {year} {2013})}\BibitemShut
  {NoStop}%
\bibitem [{\citenamefont {Efetov}(1982)}]{Efetov82}%
  \BibitemOpen
  \bibfield  {author} {\bibinfo {author} {\bibfnamefont {K.}~\bibnamefont
  {Efetov}},\ }\href {http://www.jetp.ac.ru/cgi-bin/e/index/e/55/3/p514?a=list}
  {\bibfield  {journal} {\bibinfo  {journal} {Sov. Phys. JETP}\ }\textbf
  {\bibinfo {volume} {55}},\ \bibinfo {pages} {514} (\bibinfo {year}
  {1982})}\BibitemShut {NoStop}%
\bibitem [{\citenamefont {Efetov}(1983)}]{Efetov83}%
  \BibitemOpen
  \bibfield  {author} {\bibinfo {author} {\bibfnamefont {K.}~\bibnamefont
  {Efetov}},\ }\href {\doibase 10.1080/00018738300101531} {\bibfield  {journal}
  {\bibinfo  {journal} {Advances in Physics}\ }\textbf {\bibinfo {volume}
  {32}},\ \bibinfo {pages} {53} (\bibinfo {year} {1983})}\BibitemShut {NoStop}%
\end{thebibliography}%

%\end{thebibliography}

\end{document}